\documentclass[aps,amsfonts,amsmath,prd,preprint,nofootinbib]{revtex4}
\pdfoutput=1
\newcommand{\beq}{\begin{equation}}
\newcommand{\eeq}{\end{equation}}
\usepackage{graphicx}
\usepackage{natbib}

\usepackage{hyperref}

\usepackage{amsmath}
\DeclareMathOperator{\sech}{sech}
\newcommand{\per}{{\rm pert}}

\begin{document}

\title{Exciting the Domain Wall Soliton }

\author{Jose J. Blanco-Pillado{$^{1,2,}$} \footnote{josejuan.blanco@ehu.eus}, Daniel Jim\'enez-Aguilar{$^{1,}$} \footnote{daniel.jimenez@ehu.eus} and Jon Urrestilla{$^{1,}$} \footnote{jon.urrestilla@ehu.eus}}


\affiliation{ $^1$ Department of Theoretical Physics, UPV/EHU, 48080, Bilbao, Spain \\
$^2$ IKERBASQUE, Basque Foundation for Science, 48011, Bilbao, Spain \\
}

 

\begin{abstract}
Many solitonic configurations in field theory have localized bound states
in their spectrum of linear perturbations. This opens up the possibility of 
having long lived excitations of these solitons that could affect their
dynamics. We start the study of these effects in the simplest configuration
of a domain wall kink solution in the $\lambda \phi^4$ theory in $1+1$
dimensions. We show that this solution has a single bound state and numerically 
study its slow decay rate in flat space. We then investigate the amplitude of this excitation by simulating a cosmological 
phase transition that leads to the formation of these kinks in an expanding 
universe. We find that kinks get formed with a $20\%$ excess of energy with 
respect to their lowest energy configuration. We also explore the kink solution
interacting with a thermal bath and extract the amplitude of the localized excitation as a
function of temperature. We note that this amplitude increases with temperature
but again the extra energy in the kink never goes over the $20\%$ level.
Finally, we argue that this extra energy may have important consequences
in the subsequent evolution of defects in numerical simulations. 

\end{abstract}

\maketitle

\section{Introduction}
 
Non linear field theories are present in many interesting physical situations. The presence of these
non-linearities allow these theories to have, in their spectra of excitations, solitonic like configurations that
could play a significant role in their dynamics. Examples of this type of solutions can be found in 
many areas of physics from condensed matter \cite{Bishop80} to particle physics \cite{Skyrme:1961vq}, 
string theory \cite{Polchinski:1998rq}
or cosmology \cite{Vilenkin:2000jqa}. 

These solitonic solutions can be obtained by considering the lowest energy configurations of the
fields involved with a particular boundary condition at infinity. An example of this idea can be found in 
theories with topological defects where the stability of the solutions is guaranteed by their topological
charge. These objects have been extensively studied in the literature in the past decades. In $3+1$
dimensions one can find solitonic objects of different dimensionality like domain walls \cite{Zeldovich:1974uw},
cosmic strings \cite{Nielsen:1973cs} or monopoles \cite{tHooft:1974kcl,Polyakov:1974ek}\footnote{One could 
also consider instantons and textures as part of this family of solutions. However, we do not have much to add about these
type of objects in this paper.}.

On the other hand, solitons are not the only long lived localized structures that can exist in non-linear
field theories. Many of the same models that have solitons also support lump-like field arrangements
that could have extremely long lifetimes. One of the most prominent examples of this type of objects are 
oscillons\footnote{Depending on the theory they are also called Quasi-breathers or Pulsons.} 
\cite{Kudryavtsev:1975dj,Bogolyubsky:1976nx,Bogolyubsky:1976yu,Gleiser:1993pt}. These configurations 
are not exact solutions of the equations of motion and present a very slow radiating process that allows the 
energy to leak to infinity in terms of small amplitude oscillations of the field around its vacuum. 

In this paper we will show how in some field theories it is natural to expect the existence
of some excited soliton solutions whose properties resemble in many cases the oscillon-like
objects. In fact, we will show that the interactions between solitons and oscillons would
give rise to these excited states which decay with a characteristically long time scale.
These localized excitations could become relevant for the dynamics of the solitons, specially if they
last for long periods of time. If this was the case, they could become important
even at low energies. This would significantly modify the standard 
lore where at energies well below the energy scale of the soliton one expects that its internal deformations
 would not be excited. 
 
 Usually, one assumes that at low enough energies solitons should behave as coherent objects of their 
 dimensionality, namely, walls, strings or point like objects. In this limit, one can easily write an effective 
 theory for their low energy dynamics: the so-called thin wall approximation, where the soliton's width 
 does not play any role. In the case of relativistic strings, this type of dynamics is described by the 
 Nambu-Goto (NG) action \cite{Olum:1998ag,Olum:1999sg}. Lattice field theory simulations of individual strings have shown 
 that this NG action is indeed a very good approximation for the motion of the strings. In fact, 
 simulations show that only regions of the string evolution that reach high curvature deviate from the 
 NG predictions \cite{Matsunami:2019fss}.

However, large scale lattice field theory simulations of cosmic strings seem to imply that a string network behaves
quite differently from what is predicted by the NG dynamics \cite{Vincent:1997cx,Hindmarsh:2017qff}. These 
differences are quite substantial and affect very significantly the observational effects on strings, in 
particular their predictions regarding their gravitational wave signatures 
\cite{Blanco-Pillado:2017rnf,Auclair:2019wcv,Hindmarsh:2017qff}. Given the significant progress in the observational 
bounds on gravitational wave signals, this uncertainty in the predictions from cosmic string 
networks is becoming a very pressing issue that needs to be resolved. One of the suggestions that 
has been put forward in the literature in trying to explain the different behavior of lattice field theory 
strings is the possibility that cosmological evolution of the network could lead to very excited 
strings \cite{Hindmarsh:2017qff}. These excitations could affect the evolution of the strings if 
they are present for long enough in the history of the network.

In this paper we take a step back and study the evolution of a much simpler field theory model based on a
single real scalar field, the $\lambda \phi^4$ model. We will argue that this simple field theory has many
of the properties that are important to understand the cosmic string scenario. The theory possesses field
theory solitonic solutions that can be in an excited state. In the following sections we will show that 
these excited states have a very long lifetime that can affect the long term
evolution of their field theory network simulations. However, in order to understand the impact of these excited
modes, one should first show that they can be dynamically excited. In this paper we
start this investigation by looking at the possible excitation of the solitons in this
model through their cosmological formation in a lattice field theory model in 
$1+1$ dimensions.

The organization of the paper is the following. We introduce in Section (\ref{the-kink-solution}) the exact solution
for the kink soliton that we will be discussing in the rest of the paper. In Section (\ref{perturbations}) we describe
the spectrum of excitations of this soliton and their physical interpretation. In Section (\ref{the-breather}) we
comment on the possible existence of breather solutions in the $\lambda \phi^4$ theory that
we are studying. In Section (\ref{decay-flat-space}) we study numerically the non-linear decay of the bound state
modes in flat spacetime. In Section (\ref{evolution-expanding-background}) we investigate the solutions 
in an expanding universe. In Section (\ref{phase-transition}) we simulate the formation of these kinks in a phase transition 
in an expanding universe. We can then obtain the average level of excitation of the kinks at 
formation. In Section (\ref{thermal-bath}) we also explore the process of excitation when the kinks are in contact with a thermal
bath and look at the dependence of the excitation with the background temperature. We conclude with a 
brief discussion of the possible impact of these results.
In an attempt to make the paper easier to read, we have moved many of the technical
discussions to the appendices. \\\\
The animations corresponding to some of the simulations described in the following sections can be found at \url{http://tp.lc.ehu.es/earlyuniverse/kink-simulations/}.

\section{The kink solution in $\lambda \phi^4$}
\label{the-kink-solution}

Let us introduce the model we are interested in, the so-called $\lambda \phi^4$ model, whose
action is given by
\beq
S_{\lambda \phi^4} = \int{dx^2 \left[\frac{1}{2} \partial_{\mu} \phi  \partial^{\mu} \phi -V(\phi)\right]} =  \int{dx^2 \left[\frac{1}{2} \partial_{\mu} \phi  \partial^{\mu} \phi -\frac{\lambda}{4}\left(\phi^2 - \eta^2\right)^2\right]}.
\eeq

This is a 1+1 dimensional model where we are using a metric signature of (+1,-1). 
The potential of this model has two degenerate minima at $\phi = \pm \eta$ and the
fluctuations around these minima are characterized by perturbative excitations
of mass  $m^2 = 2\lambda \eta^2$. 

Apart from these excitations, it is also
well known that this theory possesses  non-perturbative states that interpolate 
between both vacua. For example, the so-called kink solutions interpolate between 
the $\phi = - \eta$  and the $\phi = + \eta$ minima as $x$ grows. Due to translational 
invariance, the zero of the kink solution is a free parameter. The kink solution  
centered at the point $x_0$ is given by \cite{Rajaraman:1982is}
\beq
\phi_{k,x_0}(x) = \eta \tanh \left(\sqrt{\frac{\lambda}{2}} \eta (x-x_0) \right)=  \eta \tanh \left( \frac{m}{2} (x-x_0)\right)~.
\eeq
We will be mostly dealing with the kink centered at the origin ($x_0=0$): $\phi_k\equiv\phi_{k,x_0=0}$.
 
These solitonic \footnote{Strictly speaking these are not solitons, see below.} solutions have an energy density which is 
exponentially localized around their center, and one can estimate their
width to be of the order of $\delta_K \sim \sqrt{\frac{2}{\lambda}} \eta^{-1}$.

Furthermore, the total energy of these states can easily be computed to be
\beq
M_k= \frac{2 \sqrt{2 \lambda}}{3} \eta^3~~.
\label{mass-kink}
\eeq

One can also find the solutions that interpolate between these vacua but with
the opposite boundary conditions at infinity, in other words, solutions with a different
orientation, which are normally described as anti-kink solutions. We will refer to both kinks 
and anti-kinks generically as kinks, unless the distinction is relevant.

Many different aspects of kink and anti-kinks and their interactions have been 
extensively studied in the literature over the years using a combination
of analytic as well as numerical techniques \cite{Vachaspati:2006zz}. One interesting point about this
model is its lack of integrability. This makes this model quite different in many
respects to its integrable counterpart, the Sine-Gordon (SG) model \cite{Dashen:1975hd}.

A simple example where one can see the difference between these models
is in the interaction between kink and anti-kink. In the SG model, the two solitons
scatter off each other without any radiation. On the other hand, in the $\lambda \phi^4$ model, 
one has a complicated outcome of the collision that depends
on the initial state \cite{AKL,Campbell:1983xu,Anninos:1991un}.
Since kinks  in the $\lambda \phi^4$ model 
radiate upon their interactions, and also  since they 
suffer excitations, they are not strictly speaking solitons, and some authors refer to this
type of non-perturbative configurations as solitary waves.
We will use both these terms indistinguishably throughout the paper.

Another important case where these models are different is in the existence
of breather solutions. In the SG case, one can find analytic solutions describing
a bound state of kink and anti-kink oscillating around their center of mass \cite{Rajaraman:1982is}.
In the $\lambda \phi^4$ model, one can show that there is no stable configuration
of this form \cite{Segur:1987mg}. This does not mean, however, that there are not 
long lasting configurations of this type. In fact, one can easily create these
localized oscillating states that slowly decay by emitting radiation.
Furthermore, their interpretation as kink-anti-kink bound states 
was also given after some of the first numerical experiments
performed in this model \cite{Kudryavtsev:1975dj}. We will discuss more about 
these type of solutions later in the paper in relation to the kink excitations.

\section{The spectrum of excitations around the kink}
\label{perturbations}

Let us  characterize the small perturbations about the static
kink solutions presented earlier. We will assume
the field to be separated into the kink solution plus perturbations as
\beq
\phi(x,t) = \phi_k(x) + \psi (x,t) \,,
\eeq
where  $|\psi| << \eta$. The linearized equations of
motion for these perturbations become
\beq
\ddot \psi - \psi'' + \lambda \left[ 3\phi_k^2(x) -\eta^2\right] \psi = 0~,
\eeq
where the dots and primes denote derivatives with respect to time and space
respectively. Taking an oscillatory ansatz for the perturbations of the form
 $\psi(x,t) \propto e^{-iwt} f(x)$, the equation for the transverse profile of the 
 perturbations has the form of a Schr\"odinger-like equation:
\beq
-f''(x) + U(x) f(x) = w^2 f(x)~,
\eeq
 with  potential 
\beq
U(x) = \lambda \left[ 3\phi_k^2(x) -\eta^2\right] ~.
\eeq

This  turns out to be a completely
solvable potential  (see \cite{MorseandFeshbach}). Its spectrum is composed of two discrete
modes and a continuum of scattering states \cite{Rajaraman:1982is}. The two discrete modes are
\beq
f_0 (x) = \sech^{2} \left( \frac{m x} {2}\right) ~~~~~~~\text{with}~~~~~~ w_0 = 0~,
\eeq
and

\beq
f_s (x) = \sinh \left( \frac{m x} {2}\right) ~ \sech^{2} \left( \frac{m x} {2}\right) ~~~~~~~\text{with}~~~~~~ w_s = \frac{\sqrt{3}}{2}m~,
\label{eq:bound-state-mode}
\eeq
and the continuum of scattering states have a functional form given by
\beq
f_k(x) = e^{ikx} \left[3 \tanh^2 \left( \frac{m x} {2}\right) - 1- \frac{4 k^2}{m^2} -i~ \frac{6 k}{m}~ \tanh\left( \frac{m x} {2}\right) \right]~,
\label{eq:propagating}
\eeq
where $w_k^2 = k^2 + m^2$. Thus, their frequencies are in the range
$m < w_k < \infty$.  These last functions become the plane wave
solutions for  asymptotically large values of $x$. These plane waves are the ones associated with
 the continuum of perturbative fluctuations around the vacuum, the
asymptotic particle states.  

The physical interpretation of the two discrete modes is  straightforward. The zero mode
describes small rigid perturbations of the position of the soliton itself. One can see this by 
computing the change of the field distribution due to a small shift in the kink position, namely
\beq
\phi_k(x+ \delta x) \approx \phi_k(x) + \frac{d \phi_k(x)}{dx} \delta x , \
\eeq
and noticing that
\beq
 \frac{d \phi_k(x)}{dx} \delta x \propto \sech^{2} \left( \frac{m x} {2}\right) \propto f_0(x)~.
\eeq

The other bound state modifies the width of the kink. One can also build some 
intuition for the spatial shape of this mode by performing a small variation of the thickness
of the kink solution to get
\beq
\phi_k(x/(1+\delta \Delta)) \approx \phi_k(x) - x \frac{d \phi_k(x)}{dx} \delta \Delta  \approx 
\phi_k(x) - \frac{m\eta}{2} x \sech^{2} \left( \frac{m x} {2}\right) \delta \Delta \,.
\eeq

Direct comparison between this last expression and  the bound state $f_s(x)$ shows that they 
have a  very similar profile, which suggests the name  {\it shape mode}\footnote{A collective
coordinate approach can also be used to single out the translation as well as the width
degrees of freedom. It is interesting to see that this approach yields a spectrum quite similar to the
linear field theory calculation. (See for example \cite{1983PhRvB..28.3587R}).} for this
bound state. Throughout this paper we will refer to this state as {\it shape mode}, {\it bound state} or {\it internal mode}.

\section{Breather solutions}
\label{the-breather}

As we mentioned in the introduction, this model has other type of interesting 
configurations with an extraordinarily long lifetime,
first discovered by numerical experiments a long time ago in \cite{Kudryavtsev:1975dj}. They describe 
the oscillation of the kink and anti-kink around each other and they share many 
of the properties of the Sine-Gordon breather exact solutions \cite{Rajaraman:1982is}. This motivated the
pursue of approximate expressions for these type of time-dependent configurations as
well as their numerical exploration by several groups \cite{Campbell:1983xu}.
In \cite{Dashen:1975hd}, the authors gave an approximate ansatz for these objects
as a perturbative expansion in the (small) amplitude. However, it was
shown in \cite{Segur:1987mg} that all these attempts to construct these periodic, localized
solutions were flawed by the presence of a radiating tail\footnote{This is why many people
refer to these configurations as quasi-breathers to remark their lack of true periodicity.}. This implied that
all these oscillating solutions had a finite lifetime, although in some cases
they could stay around for a  long enough time, such that they could play a significant
role in the dynamics. One can think of these configurations
as the $1+1$ dimensional version of the pulson \cite{Bogolyubsky:1976nx,Bogolyubsky:1976yu} 
or the oscillon \cite{Gleiser:1993pt} 
that appear in many higher dimensional field theories.

Since we will later on encounter these objects in the course of our simulations,  it 
is useful to  describe  them now. In the small amplitude
regime, one can find solutions of the equations of motion to order $O(\tilde\epsilon^2)$ of the form \cite{Dashen:1975hd}
\beq
\phi_B(x,t) = \eta \left(1 + \frac{2}{3}~ \tilde \epsilon ~\sech (\tilde \epsilon \sqrt{\lambda} \eta x) \cos(\tilde \omega_B \sqrt{\lambda} \eta  t) + \tilde \epsilon^2 ~ \sech (\tilde \epsilon \sqrt{\lambda} \eta x)^2 \left(  \frac{1}{6} \cos( 2\tilde \omega_B \sqrt{\lambda} \eta  t) - \frac{1}{2}  \right)\right)\,, \nonumber
\eeq
where one must adjust the frequency of oscillation for different amplitudes and 
widths of the object using the relation
\beq
\tilde \omega_B = \sqrt{2 - \tilde \epsilon^2}~~~.
\eeq
 
Looking at this expression, one realizes that, at the lowest order, these solutions
always oscillate with frequencies lower than the mass of the perturbative
excitations in the vacuum.  In other words,  we always have $w_B = \tilde \omega_B \sqrt{\lambda} \eta < m$.
This means that, at the lowest order, these excitations are not 
able to emit radiation (see the discussion on scattering states in the previous section). 
However, it is clear that  $O(\tilde \epsilon^2)$ terms 
or higher could decay by coupling these solutions to the scattering modes.
This is why these configurations have such a slow decay rate and last for
such a long time.

In the following we will show that, in a high energy background,  these objects can be 
created  together with kinks, and  thus they could easily run into each other. 
In fact, in the course of our simulations we have seen breathers interacting with kinks
and exciting the kink's shape mode in a significant way. This is easily understood  by 
noticing that the breather like solutions  have a large overlap with the shape 
mode. This relation can be made much more apparent by looking at a particular type
of breather with amplitude $\tilde \epsilon = \frac{1}{\sqrt{2}}$, such
that its width and  fundamental frequency agree with the characteristic
ones of the shape mode. In this case, at lowest order, the breather looks like
\beq
\phi_B(x,t) = \eta \left(1 + \frac{2}{3\sqrt{2}}~\sech \left(\frac{m x}{2}\right)  \cos\left(\frac{\sqrt{3} m}{2}  t \right) + O(\tilde \epsilon^2)\right),
\eeq
which suggests that an excited kink with a non zero amplitude of its shape mode can be understood
as the combination of the kink and the breather. In other words, we notice that
\beq
\left[ \tanh \left( \frac{m}{2} x\right)~\times \phi_B(x,t) \right] -\phi_k(x) \propto f_s (x)\cos(w_s t)\,.
\eeq

This realization suggests that  the localized shape mode of the kink can be thought of 
as a sort of breather trapped by the soliton. The connection between these types of objects
has also been explored in the literature  recently in \cite{Romanczukiewicz:2018gxb}.

\section{Numerical Investigations for the shape mode}
\label{decay-flat-space}

\subsection{Preliminaries}

Since we will make extensive use of numerical simulations throughout the paper,  it will 
be convenient to reformulate our theory in terms of the following dimensionless quantities:
\beq
\tilde \phi = \frac{\phi}{\eta} ~~~~~~~~~~~~~~ \tilde x = \sqrt{\lambda} \eta x~~~~~~~~~~~~~~\tilde t = \sqrt{\lambda} \eta t~~.
\eeq
With these redefinitions, the action becomes
\beq
S = \eta^2 \int{d^2 \tilde x ~\left[\frac{1}{2} \partial_{\mu} \tilde \phi ~ \partial^{\mu} \tilde \phi - \frac{1}{4} \left( \tilde \phi^2 - 1\right)^2\right] }~,
\eeq
and the equation of motion in terms of the rescaled quantities is
\beq
\ddot {\tilde \phi} - \tilde \phi'' - \tilde \phi + \tilde \phi^3 = 0~.
\label{eomnum}
\eeq

Note that in these units the mass of perturbative excitations around the vacua is $\tilde m=\sqrt{2}$, and, for example, the frequency of the 
shape mode is $\tilde w_s=\sqrt{\frac{3}{2}}$. The dimensionless period for the shape mode is 
then $P=2\pi/\tilde w_s\approx5.1302$, which will be used as a unit of time in most of the subsequent plots. In the rest 
of the paper we only work with dimensionless quantities unless otherwise specified, but for simplicity
of the notation we will drop the tildes over them.

The solutions of these equations will describe the evolution of the physical system in any point
of the two dimensional space of field theories parametrized by $(\eta, \lambda)$. Note that even though 
the parameter $\eta^2$ is absent from the dimensionless classical equations of motion, it appears
as an overall coefficient in the action, and therefore  it does have implications at the quantum level\footnote{In fact,
$\eta$ plays the role of the inverse of a coupling constant so the weak coupling regime corresponds
to the $\eta >> 1$ limit. One can see this by looking at the ratio of the soliton and elementary excitation
masses, namely $M_k / m \sim \eta^2$.}. This will be of importance in Sections~\ref{phase-transition}
and \ref{thermal-bath}.

\subsubsection{Extracting the amplitude of the shape mode}

In the following we will describe several different lattice simulations that we have
done to understand the level of excitation that the kink solutions can 
acquire. It is therefore paramount for us to be able to quantify this in 
a precise way. We do this is by computing the amplitude of the shape
mode of the kink directly from the simulation data.

We first note that the solutions to the Schr\"odinger-like equations 
for the linear perturbations described in the previous section  form an 
orthonormal basis. Thus, the general expansion for a linear perturbation around the kink can be written in 
dimensionless units as
\beq
\delta (x, t) = \hat A_0 \bar f_0(x) + \hat A_s \bar f_s( x) \cos(w_s  t)+  \int  {\rm d} k \hat A_k Re[\bar f_k( x)  e^{-i w_k  t}]~,
\eeq
where we have denoted by $\bar f_i(x)$ the normalized mode functions \footnote{The normalized
shape mode function in dimensionless units is given by 
\beq
\bar f_s (x) =\frac{\sqrt{3\sqrt{2}}}{2}\sinh \left( \frac{x} {\sqrt{2}}\right) ~ \sech^{2} \left( \frac{x} {\sqrt{2}}\right) 
\eeq
}.

However, since the model is non-linear, the interaction between different modes will 
make the $\hat A_i$ coefficients above time-dependent:
\begin{eqnarray}
\delta (x,t) &=& \hat A_0(t) \bar f_0(x) + \hat A_s(t) \bar f_s(x) \cos(w_s t)+ \int  {\rm d}k \hat A_k(t) Re[\bar f_k(x)  e^{-i w_k t}]\\&=&A_0(t) \bar f_0(x) + A_s(t) \bar f_s(x)+ f_r(x,t)\,,
\label{hats}
\end{eqnarray}
where (in the last line) we have absorbed all the time dependency into the $A_i (t)$ and we have defined  
the integral carrying the information for the radiation as  $f_r$. 

To extract the values of $ A_i$ (in particular, $ A_s(t) $) given a configuration $\phi(x,t)$ 
consisting of a kink plus excitation, we first obtain the point $x_0$ where the field $\phi$ 
goes through zero, and define that as the center of the kink. We then calculate the 
perturbations around the kink as
\beq
\phi_\per (x,t)= \phi (x,t) - \phi_{k,x_0}(x,t)
\label{pert}
\eeq
and finally project the perturbations over the shape mode by computing:
\beq
A_s(t) = \int_{-L/2}^{L/2} {dx \, \phi_\per (x,t) ~\bar f_s (x-{x_0})}~.
\label{numerical-amplitude}
\eeq

This is the quantity that we will follow during the evolution of the kink
in different situations and that we will compare with analytic predictions
in the subsequent sections.

\subsubsection{Numerical calculations in the lattice}

Throughout this work we have solved the equation of motion (\ref{eomnum}) in a lattice. The 
details of its implementation can be found in Appendix~(\ref{numericaldetails}), but here
we would like to emphasize a few points that will become important in the rest of the paper. 
Since we are dealing with a $1+1$ dimensional lattice, we can use a large array of points 
in the spatial direction without too much computational cost.  Moreover, we have written
the code in a parallelized fashion so we can use many nodes to implement
the evolution of a large volume. The combination of these two facts has allowed
us to explore a considerable large volume in our simulations while still faithfully representing
the dynamics of the fields. 

This will become more advantageous later in the paper, since some of our simulations 
will be performed in an expanding background with comoving coordinates. In that 
situation, there is a well-known problem:
the comoving size of the solitons in these expanding 
backgrounds shrinks with time. This means that by the end of the simulation 
one could have too small a number of comoving lattice points in the relevant central 
region of the kink. That is why the use of a large number of points, and a parallel code, 
are helpful to make sure that our final
configurations had at least $20$ points covering the important central region
where the bound state has its support. 

A somewhat popular way to deal with this issue is the so-called "fat string algorithms" \cite{Press:1989yh},
where the equations of motion are modified to change the rate of contraction
of the soliton width in comoving coordinates. This allows the possibility to 
track down the position of the solitons without introducing more points in the lattice.
However, this method affects, by construction, the physical width of the soliton in the
simulation. This will clearly distort the level of excitation of the shape
mode in an artificial way, hence we refrain from using such algorithms here.

It is also worth  emphazising  that we use absorbing boundary conditions \cite{ABC} in our simulations. As 
their name indicates, these boundary conditions work in such a way that they mimic the 
absorption of waves by the boundary.  As we will see shortly, we will need to simulate the 
system for times much, much longer than the light-crossing time of the box. 
This poses a numerical problem since we do not want to have the radiation energy bounce back from the edge of the
simulation. This is a particularly important  concern in this case because in one spatial dimension 
there will not be any dilution of the radiation. That is why we use absorbing boundary conditions, since 
they allow us to  run the simulation for as long as we want and compare the
results with the theoretical predictions. Actually,  since we will be mostly dealing with frequencies 
coming from the radiation from the shape mode, we have tuned these boundary conditions such 
that they will be most effective at those frequencies (See the discussion in the Appendix \ref{AbsorbingBC}.).

\subsection{Lifetime of the excitation}
\label{lifetime}

Our starting point is a detailed study of the shape mode excitation of the kink and its decay. In order to do that, we
initialize our ($1+1$) dimensional lattice field with a kink at the center of the box and we add to it a small perturbation
of the form of the linear excitation described earlier as the shape bound state:
\beq
\phi(x,t) = \phi_k(x) + A(0) \times \bar f_s(x)\,.
\eeq

We are taking a small amplitude, $A(0) < 1$, and we do not give the field any initial velocity. 
Following the discussion on linear modes given earlier, one would
think that this configuration should stay oscillating without any variation. The reason for this is 
that the value of its frequency is smaller than the ones that are allowed to propagate outside of 
the kink towards infinity. However, the full system is non-linear, so the amplitude of this bound 
state is expected to decrease over time by emitting radiation at a small
rate.

At the lowest order, the non-linear terms will produce a radiation field with a frequency which doubles 
the bound state one and a quadratically suppressed amplitude. This clearly means that the
lifetime of these perturbations will be much, much longer than the typical scale of the problem,
i.e., the light-crossing time of the width of the soliton. Actually, it is likely to be a much longer time
than the light-crossing time of the simulation box as well. This is why we use the aforementioned 
absorbing boundary conditions (see Appendix~\ref{AbsorbingBC}).

\begin{figure}[h!]
\includegraphics[width=16cm]{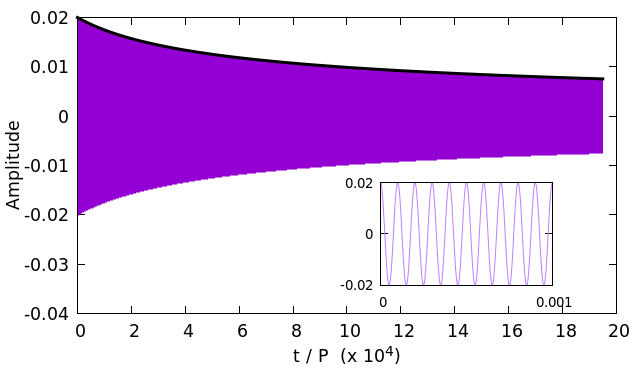}
\caption{Amplitude of the shape mode $A(t)$ as a function of time (displayed as number 
of periods or oscillations). $P$ is the dimensionless period of the shape mode bound state, $P=2\pi/\tilde w_s\approx5.1302$. The black 
curve corresponds to the analytical second order approximation to $\hat A(t)$, namely 
equation (\ref{second order theoretical curve}) with $A (0) = \hat A(0) =0.02$. Also shown 
is a zoom of the first oscillations of the amplitude of the internal mode; $A(t)$.}
\label{fig:amplitude comparison}
\end{figure}

The problem of studying the decay of a small fluctuation like the one we just presented 
has been analyzed by Manton and Merabet \cite{Manton:1996ex} using analytical 
techniques. Their conclusion was that the amplitude of the
shape mode should decrease with time following the expression\footnote{We give in the 
Appendix~ \ref{analytic-decay}   a quick derivation of this expression following the calculations   
in \cite{Manton:1996ex} and adapting them to our current notation.}
\beq
\frac{3}{4} \frac{d\hat A^2}{dt} = - 0.0112909 ~\hat A^4\,,
\eeq
which means that the amplitude should have the following time dependence:
\beq
\label{second order theoretical curve}
\hat A(t)^{-2} = \hat A(0)^{-2} + 0.0150546 ~ t ~.
\eeq

Using the results in \cite{Manton:1996ex} we can also obtain the predicted form
of the radiation field to be
\begin{equation}
f_{r}\left(x,t\right)=\frac{3\sqrt{3}\,\pi \hat A^{2}}{8\sinh\left(\pi\sqrt{2}\right)}\cos\left[2\sqrt{\frac{3}{2}}\,t-2x-\arctan\left(\sqrt{2}\right)\right]\,.
\label{second order eleven}
\end{equation}

We have run different simulations starting with several values of the 
parameter $A(0)$ in order to check the accuracy of these theoretical expectations.
In each case we extracted the value of the amplitude from the simulation by
finding the perturbations around the kink and then projecting the perturbations over 
the shape mode (see Eq. (\ref{numerical-amplitude})).

Our simulations demonstrate that the analytic predictions work perfectly in the case of 
a small initial amplitude for the bound state. In Fig~\ref{fig:amplitude comparison} we show
the comparison between  the numerically time dependent amplitude $A(t)$ extracted using
the expression given in Eq.  (\ref{numerical-amplitude}) and the analytic
prediction for $\hat A(t)$ (\ref{second order theoretical curve}). Remember that  $A(t)$ 
carries information about the oscillatory part of the amplitude, whereas $\hat A(t)$ follows the 
envelope created by the maxima of the oscillations. Note that we have evolved this numerical 
computation for $2\times 10^5$ times the period of the oscillation of the bound state.

\begin{figure}[h!]
\begingroup
  \makeatletter
  \providecommand\color[2][]{%
    \GenericError{(gnuplot) \space\space\space\@spaces}{%
      Package color not loaded in conjunction with
      terminal option `colourtext'%
    }{See the gnuplot documentation for explanation.%
    }{Either use 'blacktext' in gnuplot or load the package
      color.sty in LaTeX.}%
    \renewcommand\color[2][]{}%
  }%
  \providecommand\includegraphics[2][]{%
    \GenericError{(gnuplot) \space\space\space\@spaces}{%
      Package graphicx or graphics not loaded%
    }{See the gnuplot documentation for explanation.%
    }{The gnuplot epslatex terminal needs graphicx.sty or graphics.sty.}%
    \renewcommand\includegraphics[2][]{}%
  }%
  \providecommand\rotatebox[2]{#2}%
  \@ifundefined{ifGPcolor}{%
    \newif\ifGPcolor
    \GPcolorfalse
  }{}%
  \@ifundefined{ifGPblacktext}{%
    \newif\ifGPblacktext
    \GPblacktexttrue
  }{}%
  \let\gplgaddtomacro\g@addto@macro
  \gdef\gplbacktext{}%
  \gdef\gplfronttext{}%
  \makeatother
  \ifGPblacktext
    \def\colorrgb#1{}%
    \def\colorgray#1{}%
  \else
    \ifGPcolor
      \def\colorrgb#1{\color[rgb]{#1}}%
      \def\colorgray#1{\color[gray]{#1}}%
      \expandafter\def\csname LTw\endcsname{\color{white}}%
      \expandafter\def\csname LTb\endcsname{\color{black}}%
      \expandafter\def\csname LTa\endcsname{\color{black}}%
      \expandafter\def\csname LT0\endcsname{\color[rgb]{1,0,0}}%
      \expandafter\def\csname LT1\endcsname{\color[rgb]{0,1,0}}%
      \expandafter\def\csname LT2\endcsname{\color[rgb]{0,0,1}}%
      \expandafter\def\csname LT3\endcsname{\color[rgb]{1,0,1}}%
      \expandafter\def\csname LT4\endcsname{\color[rgb]{0,1,1}}%
      \expandafter\def\csname LT5\endcsname{\color[rgb]{1,1,0}}%
      \expandafter\def\csname LT6\endcsname{\color[rgb]{0,0,0}}%
      \expandafter\def\csname LT7\endcsname{\color[rgb]{1,0.3,0}}%
      \expandafter\def\csname LT8\endcsname{\color[rgb]{0.5,0.5,0.5}}%
    \else
      \def\colorrgb#1{\color{black}}%
      \def\colorgray#1{\color[gray]{#1}}%
      \expandafter\def\csname LTw\endcsname{\color{white}}%
      \expandafter\def\csname LTb\endcsname{\color{black}}%
      \expandafter\def\csname LTa\endcsname{\color{black}}%
      \expandafter\def\csname LT0\endcsname{\color{black}}%
      \expandafter\def\csname LT1\endcsname{\color{black}}%
      \expandafter\def\csname LT2\endcsname{\color{black}}%
      \expandafter\def\csname LT3\endcsname{\color{black}}%
      \expandafter\def\csname LT4\endcsname{\color{black}}%
      \expandafter\def\csname LT5\endcsname{\color{black}}%
      \expandafter\def\csname LT6\endcsname{\color{black}}%
      \expandafter\def\csname LT7\endcsname{\color{black}}%
      \expandafter\def\csname LT8\endcsname{\color{black}}%
    \fi
  \fi
    \setlength{\unitlength}{0.0500bp}%
    \ifx\gptboxheight\undefined%
      \newlength{\gptboxheight}%
      \newlength{\gptboxwidth}%
      \newsavebox{\gptboxtext}%
    \fi%
    \setlength{\fboxrule}{0.5pt}%
    \setlength{\fboxsep}{1pt}%
\begin{picture}(9070.00,5668.00)%
    \gplgaddtomacro\gplbacktext{%
      \csname LTb\endcsname
      \put(1342,704){\makebox(0,0)[r]{\strut{}$-0.0008$}}%
      \put(1342,1297){\makebox(0,0)[r]{\strut{}$-0.0006$}}%
      \put(1342,1890){\makebox(0,0)[r]{\strut{}$-0.0004$}}%
      \put(1342,2483){\makebox(0,0)[r]{\strut{}$-0.0002$}}%
      \put(1342,3076){\makebox(0,0)[r]{\strut{}$0$}}%
      \put(1342,3668){\makebox(0,0)[r]{\strut{}$0.0002$}}%
      \put(1342,4261){\makebox(0,0)[r]{\strut{}$0.0004$}}%
      \put(1342,4854){\makebox(0,0)[r]{\strut{}$0.0006$}}%
      \put(1342,5447){\makebox(0,0)[r]{\strut{}$0.0008$}}%
      \put(2194,484){\makebox(0,0){\strut{}$-20$}}%
      \put(3634,484){\makebox(0,0){\strut{}$-10$}}%
      \put(5074,484){\makebox(0,0){\strut{}$0$}}%
      \put(6513,484){\makebox(0,0){\strut{}$10$}}%
      \put(7953,484){\makebox(0,0){\strut{}$20$}}%
    }%
    \gplgaddtomacro\gplfronttext{%
      \csname LTb\endcsname
      \put(198,3075){\rotatebox{-270}{\makebox(0,0){\Large $\phi_{pert}$}}}%
      \put(5073,154){\makebox(0,0){\Large x}}%
    }%
    \gplbacktext
    \put(0,0){\includegraphics{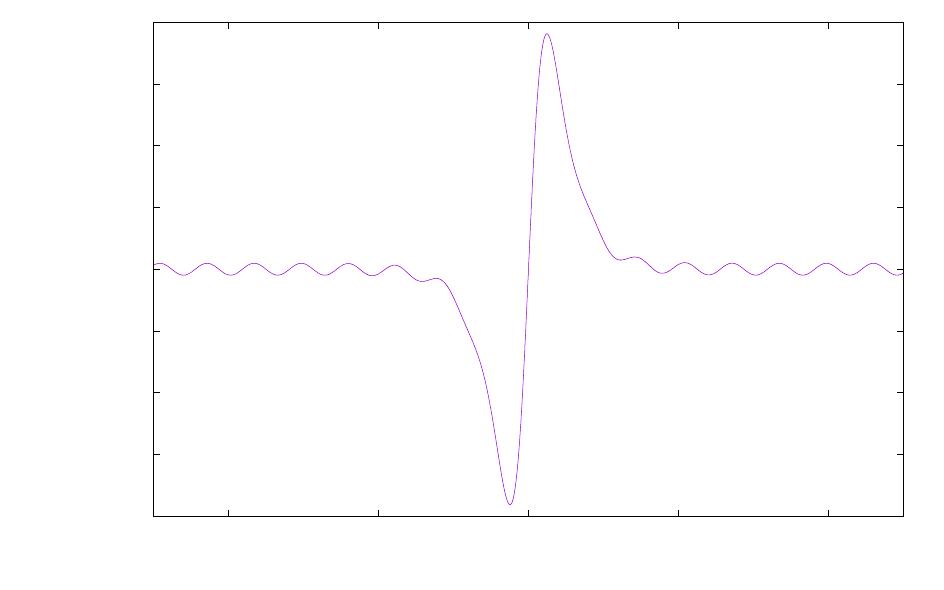}}%
    \gplfronttext
  \end{picture}%
\endgroup
\caption{Snapshot of the perturbation  $\phi_\per$ around the kink.}
\label{bound-state-and-perturbations}
\end{figure}

We also show in Fig.~\ref{bound-state-and-perturbations} a snapshot of the perturbation field 
around the kink, as defined in Eq.~(\ref{pert}). The central region can clearly be identified as 
the waveform of the shape mode (Eq.~(\ref{eq:bound-state-mode})), while the radiative small 
contribution far from the core is almost exactly a wave of twice the frequency of the shape mode  
in accordance with Eq. (\ref{second order eleven}) .\\

\begin{figure}[h!]
\includegraphics[width=16cm]{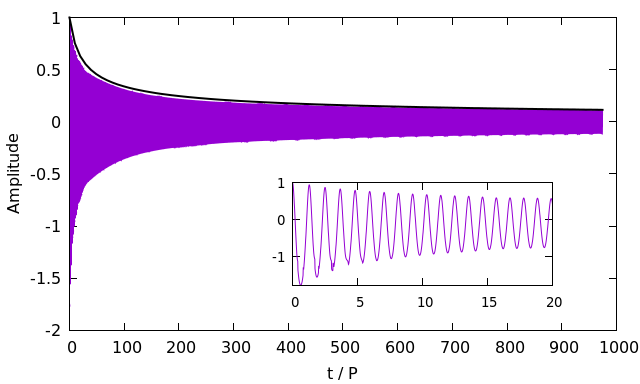}
\caption{Amplitude of the shape mode $A(t)$ as a function of time (displayed as number 
of periods or oscillations) for $A(0)=1$. We show in the inset figure the first few oscillations
of the amplitude where an initial asymmetric behaviour is clearly visible. This is a non-linear effect
due to the initial large amplitude.}
\label{fig:amplitude comparison-A_1}
\end{figure}

We have also run simulations for larger values of the initial amplitude. The agreement with
the analytic estimate is quite accurate all the way to $A(0) \sim O(1)$. This is somewhat surprising since the
amount of energy stored in the perturbation in this case is above the kink rest mass, so there
is no reason to expect this linear type behavior at this point. We show in Fig (\ref{fig:amplitude comparison-A_1}) 
the time evolution of the perturbation with $A(0) = 1$ as well as the analytic prediction
for the envelope amplitude $\hat A(t)$.

\subsection{Non-linear evolution.}

One can ask the question of what happens when one starts with a value of the
amplitude well beyond the linear regime. As we mentioned earlier, going to such high values of the amplitude
of the perturbation means that the energy of the configuration is not
 a small fraction of the energy of the kink solution. In fact, for high enough values
of $A(0)$, the energy of the initial state would be higher than the rest mass of the
energy of 3 kinks. That means that there is enough energy to create a kink pair 
leaving behind an anti-kink at the center and still have a configuration consistent with the boundary
conditions at infinity.

\begin{figure}[h!]
\begingroup
  \makeatletter
  \providecommand\color[2][]{%
    \GenericError{(gnuplot) \space\space\space\@spaces}{%
      Package color not loaded in conjunction with
      terminal option `colourtext'%
    }{See the gnuplot documentation for explanation.%
    }{Either use 'blacktext' in gnuplot or load the package
      color.sty in LaTeX.}%
    \renewcommand\color[2][]{}%
  }%
  \providecommand\includegraphics[2][]{%
    \GenericError{(gnuplot) \space\space\space\@spaces}{%
      Package graphicx or graphics not loaded%
    }{See the gnuplot documentation for explanation.%
    }{The gnuplot epslatex terminal needs graphicx.sty or graphics.sty.}%
    \renewcommand\includegraphics[2][]{}%
  }%
  \providecommand\rotatebox[2]{#2}%
  \@ifundefined{ifGPcolor}{%
    \newif\ifGPcolor
    \GPcolorfalse
  }{}%
  \@ifundefined{ifGPblacktext}{%
    \newif\ifGPblacktext
    \GPblacktexttrue
  }{}%
  \let\gplgaddtomacro\g@addto@macro
  \gdef\gplbacktext{}%
  \gdef\gplfronttext{}%
  \makeatother
  \ifGPblacktext
    \def\colorrgb#1{}%
    \def\colorgray#1{}%
  \else
    \ifGPcolor
      \def\colorrgb#1{\color[rgb]{#1}}%
      \def\colorgray#1{\color[gray]{#1}}%
      \expandafter\def\csname LTw\endcsname{\color{white}}%
      \expandafter\def\csname LTb\endcsname{\color{black}}%
      \expandafter\def\csname LTa\endcsname{\color{black}}%
      \expandafter\def\csname LT0\endcsname{\color[rgb]{1,0,0}}%
      \expandafter\def\csname LT1\endcsname{\color[rgb]{0,1,0}}%
      \expandafter\def\csname LT2\endcsname{\color[rgb]{0,0,1}}%
      \expandafter\def\csname LT3\endcsname{\color[rgb]{1,0,1}}%
      \expandafter\def\csname LT4\endcsname{\color[rgb]{0,1,1}}%
      \expandafter\def\csname LT5\endcsname{\color[rgb]{1,1,0}}%
      \expandafter\def\csname LT6\endcsname{\color[rgb]{0,0,0}}%
      \expandafter\def\csname LT7\endcsname{\color[rgb]{1,0.3,0}}%
      \expandafter\def\csname LT8\endcsname{\color[rgb]{0.5,0.5,0.5}}%
    \else
      \def\colorrgb#1{\color{black}}%
      \def\colorgray#1{\color[gray]{#1}}%
      \expandafter\def\csname LTw\endcsname{\color{white}}%
      \expandafter\def\csname LTb\endcsname{\color{black}}%
      \expandafter\def\csname LTa\endcsname{\color{black}}%
      \expandafter\def\csname LT0\endcsname{\color{black}}%
      \expandafter\def\csname LT1\endcsname{\color{black}}%
      \expandafter\def\csname LT2\endcsname{\color{black}}%
      \expandafter\def\csname LT3\endcsname{\color{black}}%
      \expandafter\def\csname LT4\endcsname{\color{black}}%
      \expandafter\def\csname LT5\endcsname{\color{black}}%
      \expandafter\def\csname LT6\endcsname{\color{black}}%
      \expandafter\def\csname LT7\endcsname{\color{black}}%
      \expandafter\def\csname LT8\endcsname{\color{black}}%
    \fi
  \fi
    \setlength{\unitlength}{0.0500bp}%
    \ifx\gptboxheight\undefined%
      \newlength{\gptboxheight}%
      \newlength{\gptboxwidth}%
      \newsavebox{\gptboxtext}%
    \fi%
    \setlength{\fboxrule}{0.5pt}%
    \setlength{\fboxsep}{1pt}%
\begin{picture}(9060.00,5660.00)%
    \gplgaddtomacro\gplbacktext{%
      \csname LTb\endcsname
      \put(747,595){\makebox(0,0)[r]{\strut{}$-1.5$}}%
      \csname LTb\endcsname
      \put(747,1408){\makebox(0,0)[r]{\strut{}$-1$}}%
      \csname LTb\endcsname
      \put(747,2221){\makebox(0,0)[r]{\strut{}$-0.5$}}%
      \csname LTb\endcsname
      \put(747,3034){\makebox(0,0)[r]{\strut{}$0$}}%
      \csname LTb\endcsname
      \put(747,3847){\makebox(0,0)[r]{\strut{}$0.5$}}%
      \csname LTb\endcsname
      \put(747,4660){\makebox(0,0)[r]{\strut{}$1$}}%
      \csname LTb\endcsname
      \put(747,5473){\makebox(0,0)[r]{\strut{}$1.5$}}%
      \csname LTb\endcsname
      \put(1639,409){\makebox(0,0){\strut{}$-40$}}%
      \csname LTb\endcsname
      \put(3220,409){\makebox(0,0){\strut{}$-20$}}%
      \csname LTb\endcsname
      \put(4801,409){\makebox(0,0){\strut{}$0$}}%
      \csname LTb\endcsname
      \put(6382,409){\makebox(0,0){\strut{}$20$}}%
      \csname LTb\endcsname
      \put(7963,409){\makebox(0,0){\strut{}$40$}}%
    }%
    \gplgaddtomacro\gplfronttext{%
      \csname LTb\endcsname
      \put(153,3034){\rotatebox{-270}{\makebox(0,0){\Large $\phi$}}}%
      \csname LTb\endcsname
      \put(4801,130){\makebox(0,0){\Large x}}%
    }%
    \gplbacktext
    \put(0,0){\includegraphics{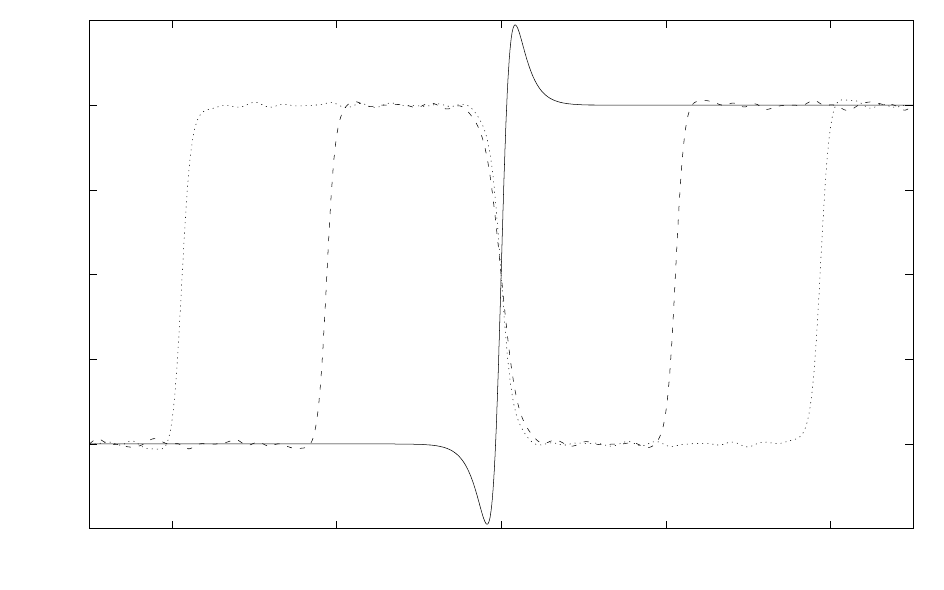}}%
    \gplfronttext
  \end{picture}%
\endgroup
\caption{Field profile at different time steps for $A (0)=1.35$. A kink-antikink-kink system forms, 
and the kinks get away from the antikink in the middle. The solid line corresponds to the initial 
state, while the dashed and dotted lines correspond to the profile at $t=80$ and $t=160$ 
respectively ($15.6$ and $31.2$ in units of the period of the shape mode $P$).}
\label{fig:kak}
\end{figure}

We have performed several simulations with those high initial energies and indeed we have
found that for $A(0) > 1.334$ the final configuration is made of a central anti-kink with a 
couple of kinks flying away to infinity in opposite directions. This clearly indicates that one can think of the
shape mode on a kink as a bound state of a kink-antikink-kink \cite{Manton:1996ex}. In fact,
this is the way in which this type of mode was first discovered by numerical
experiments in this model \cite{Getmanov:1976ik}. A collection of snapshots
of the evolution of this process is given in Fig.~\ref{fig:kak}.

Going to higher energies than the threshold of kink-antikink formation produces
a pair with higher initial velocities and leaves behind an anti-kink with some small
amplitude of the shape mode still excited.

This non-linear process shows that there is a maximum amount of energy that can 
be stored in this long lasting bound state. Beyond that, the energy is shed by the soliton
at a much faster rate in a time scale comparable to the light crossing time of the width 
of the soliton (one oscillation period). However, it is interesting to note that this 
maximum amount of energy could be even higher that the rest mass energy of the 
kink. This suggests that these modes could play a significant role in the dynamics 
of these solitons over the course of a long simulation. We shall therefore study next 
how these modes can get excited and to what extent this can happen in a realistic
setting.

\section{Numerical Investigations in an expanding background}
\label{evolution-expanding-background}

\subsection{Preliminaries}

Our main goal in this paper is to study  the  effects that excitations may have in 
the dynamics of  the cosmological evolution of solitons. As a first step, we investigate a kink 
in a dynamic spacetime. Our starting point is therefore the same
field theory as before but in a generic curved background 
with an action 
\beq
S = \eta^2 \int{d^2x \sqrt{-g} \left[ \frac{1}{2} g^{\mu \nu} \partial_{\mu} \phi  \partial_{\nu} \phi -\frac{1}{4}\left(\phi^2 - 1 \right)^2 \right]}\,,
\eeq
where $g$ denotes the determinant of the $1+1$ metric in dimensionless units given by
\beq
ds^2 = dt^2 - a^2(t)~ dx^2~.
\eeq

We will take this spacetime as a fixed background; in other words, we will disregard any
backreaction of the matter fields on this metric. Following this prescription we will specify the
form of the function $a(t)$ and explore the dependence of the
dynamics on its functional form in an attempt to simulate the
different behaviour of the cosmological spacetimes one encounters in 
$3+1$ dimensions. Our conclusions must therefore be taken as an illustration of the 
possible effects  on a full $3d$ evolution.

The equation of motion for the scalar field in this background becomes
\beq
\ddot \phi + H(t) \dot \phi - \frac{1}{a(t)^2} \phi'' + \phi \left(\phi^2 - 1\right)= 0~,
\label{eomback}
\eeq
where dots and primes denote derivatives with respect to cosmic time and comoving space
respectively, and $H(t) = \dot a(t)/a(t)$ is the Hubble rate\footnote{Note that $H$ is also written in dimensionless
units and it is related to the usual dimensional Hubble rate by the relation 
$H_{\text{physical}} = \sqrt{\lambda} \eta H$}. In a slowly expanding 
spacetime, one can approximate the soliton by a solution of fixed physical
width, namely a solution of the form

\beq
\phi(x,t) = \  \tanh \left[  a(t) \frac{x}{ \sqrt{2}}   \right]~.
\eeq

This shows that for an expanding universe the width of the soliton shrinks in comoving
coordinates. This is exactly the effect that we mentioned in the previous section. One should
therefore be able to faithfully reproduce this evolution numerically if one is interested in studying
excitations of the width of the soliton in an expanding universe.

In the following we will investigate the evolution of the kink solution in different
spacetimes and with different initial conditions in order to see whether
or not the evolution of spacetime and the environment can influence the presence
of the shape mode.

\subsection{The kink in an expanding universe}

\subsubsection{de Sitter space}

One of the simplest expanding spacetimes that we can study and that will become
useful in our future simulations is a $1+1$ dimensional de Sitter space. In this case, the
metric is given by
\beq
ds^2 = dt^2 - e^{2 H t} dx^2~.
\eeq

The equation of motion for the scalar field in our dimensionless coordinate system becomes 
\beq
\ddot \phi +  H \dot \phi - \exp{\left[-2H t\right]}\phi'' + \phi (\phi^2 -1) = 0~.
\eeq

Following the description above we can parametrize the solution of this
equation as
\beq
\phi_{dS}(x,t) = \varphi_{dS} \left[ H  a(t) x \right],
\eeq
so the function $\varphi(\xi)$ satisfies
\beq
(1 - \xi^2) \varphi_{dS}''   - 2 \xi  \varphi_{dS}' = \frac{1}{H^2}\left( \varphi_{dS}^3 - \varphi_{dS}  \right)~,
\label{eq:kink-in-dS}
\eeq
where now the primes denote derivatives with respect to the coordinate $\xi=H a(t) x$.

\begin{figure}[h!]
\begingroup
  \makeatletter
  \providecommand\color[2][]{%
    \GenericError{(gnuplot) \space\space\space\@spaces}{%
      Package color not loaded in conjunction with
      terminal option `colourtext'%
    }{See the gnuplot documentation for explanation.%
    }{Either use 'blacktext' in gnuplot or load the package
      color.sty in LaTeX.}%
    \renewcommand\color[2][]{}%
  }%
  \providecommand\includegraphics[2][]{%
    \GenericError{(gnuplot) \space\space\space\@spaces}{%
      Package graphicx or graphics not loaded%
    }{See the gnuplot documentation for explanation.%
    }{The gnuplot epslatex terminal needs graphicx.sty or graphics.sty.}%
    \renewcommand\includegraphics[2][]{}%
  }%
  \providecommand\rotatebox[2]{#2}%
  \@ifundefined{ifGPcolor}{%
    \newif\ifGPcolor
    \GPcolorfalse
  }{}%
  \@ifundefined{ifGPblacktext}{%
    \newif\ifGPblacktext
    \GPblacktexttrue
  }{}%
  \let\gplgaddtomacro\g@addto@macro
  \gdef\gplbacktext{}%
  \gdef\gplfronttext{}%
  \makeatother
  \ifGPblacktext
    \def\colorrgb#1{}%
    \def\colorgray#1{}%
  \else
    \ifGPcolor
      \def\colorrgb#1{\color[rgb]{#1}}%
      \def\colorgray#1{\color[gray]{#1}}%
      \expandafter\def\csname LTw\endcsname{\color{white}}%
      \expandafter\def\csname LTb\endcsname{\color{black}}%
      \expandafter\def\csname LTa\endcsname{\color{black}}%
      \expandafter\def\csname LT0\endcsname{\color[rgb]{1,0,0}}%
      \expandafter\def\csname LT1\endcsname{\color[rgb]{0,1,0}}%
      \expandafter\def\csname LT2\endcsname{\color[rgb]{0,0,1}}%
      \expandafter\def\csname LT3\endcsname{\color[rgb]{1,0,1}}%
      \expandafter\def\csname LT4\endcsname{\color[rgb]{0,1,1}}%
      \expandafter\def\csname LT5\endcsname{\color[rgb]{1,1,0}}%
      \expandafter\def\csname LT6\endcsname{\color[rgb]{0,0,0}}%
      \expandafter\def\csname LT7\endcsname{\color[rgb]{1,0.3,0}}%
      \expandafter\def\csname LT8\endcsname{\color[rgb]{0.5,0.5,0.5}}%
    \else
      \def\colorrgb#1{\color{black}}%
      \def\colorgray#1{\color[gray]{#1}}%
      \expandafter\def\csname LTw\endcsname{\color{white}}%
      \expandafter\def\csname LTb\endcsname{\color{black}}%
      \expandafter\def\csname LTa\endcsname{\color{black}}%
      \expandafter\def\csname LT0\endcsname{\color{black}}%
      \expandafter\def\csname LT1\endcsname{\color{black}}%
      \expandafter\def\csname LT2\endcsname{\color{black}}%
      \expandafter\def\csname LT3\endcsname{\color{black}}%
      \expandafter\def\csname LT4\endcsname{\color{black}}%
      \expandafter\def\csname LT5\endcsname{\color{black}}%
      \expandafter\def\csname LT6\endcsname{\color{black}}%
      \expandafter\def\csname LT7\endcsname{\color{black}}%
      \expandafter\def\csname LT8\endcsname{\color{black}}%
    \fi
  \fi
    \setlength{\unitlength}{0.0500bp}%
    \ifx\gptboxheight\undefined%
      \newlength{\gptboxheight}%
      \newlength{\gptboxwidth}%
      \newsavebox{\gptboxtext}%
    \fi%
    \setlength{\fboxrule}{0.5pt}%
    \setlength{\fboxsep}{1pt}%
\begin{picture}(9070.00,5668.00)%
    \gplgaddtomacro\gplbacktext{%
      \csname LTb\endcsname
      \put(946,704){\makebox(0,0)[r]{\strut{}$-1.5$}}%
      \put(946,1495){\makebox(0,0)[r]{\strut{}$-1$}}%
      \put(946,2285){\makebox(0,0)[r]{\strut{}$-0.5$}}%
      \put(946,3076){\makebox(0,0)[r]{\strut{}$0$}}%
      \put(946,3866){\makebox(0,0)[r]{\strut{}$0.5$}}%
      \put(946,4657){\makebox(0,0)[r]{\strut{}$1$}}%
      \put(946,5447){\makebox(0,0)[r]{\strut{}$1.5$}}%
      \put(1838,484){\makebox(0,0){\strut{}$-4$}}%
      \put(3357,484){\makebox(0,0){\strut{}$-2$}}%
      \put(4876,484){\makebox(0,0){\strut{}$0$}}%
      \put(6395,484){\makebox(0,0){\strut{}$2$}}%
      \put(7914,484){\makebox(0,0){\strut{}$4$}}%
    }%
    \gplgaddtomacro\gplfronttext{%
      \csname LTb\endcsname
      \put(198,3075){\rotatebox{-270}{\makebox(0,0){\Large $\phi_{dS}$}}}%
      \put(4875,154){\makebox(0,0){\Large x}}%
    }%
    \gplbacktext
    \put(0,0){\includegraphics{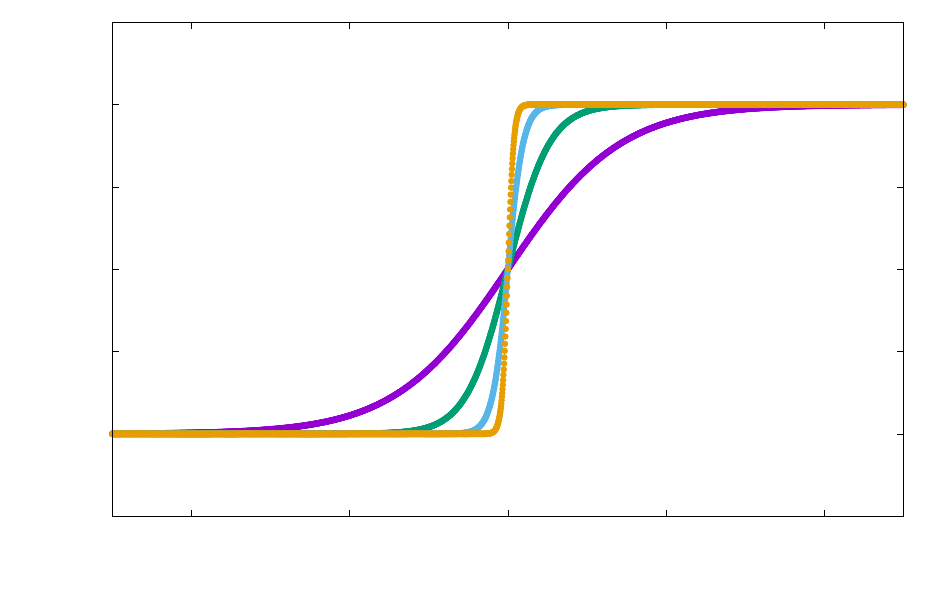}}%
    \gplfronttext
  \end{picture}%
\endgroup
\caption{Evolution of the kink solution in an expanding de Sitter space with $H=0.1$. The different 
profiles correspond to expansion factors 1, $\exp(1)$, $\exp(2)$ and $\exp(2.96)$ (purple, green, blue and orange curves respectively). The solution 
matches perfectly the one obtained from Eq.~(\ref{eq:kink-in-dS}).}
\label{fig:kink_dS}
\end{figure}

This type of equation was originally found in \cite{Basu:1993rf} in the $(3+1)d$ context. The results
here are parallel to the situation in higher dimensions since the longitudinal dimensions
to the wall do not play a significant role. We can solve this equation numerically for
different values of the Hubble parameter. For small values of $H$, the horizon distance is much larger
than the width of the kink and therefore one  expects the solution to be close
to the adiabatic solution of a kink of constant physical width.   
In Fig.~\ref{fig:kink_dS} we show the results for $H=0.1$. We observed that in this case the deviation of the
exact solution from the adiabatic one is very small. These deviations can be shown
to scale with $O(H^2)$. Furthermore, similarly to what was found in \cite{Basu:1993rf}, it can be seen 
 that there is a maximum value of $H$ beyond which there is no stationary kink solution. We will 
not explore this regime further in this paper.

We have checked that our numerical code on the lattice for a kink in a de Sitter universe
reproduces the exact scaling solution given by Eq. (\ref{eq:kink-in-dS}). We have evolved the discrete equations
of motion over several Hubble times without any visible excitation of the shape mode. See 
Fig~\ref{fig:kink_dS} for a sample of solutions in comoving space. 

Furthermore, in the regime where there is a
separation of scales between the width of the soliton and the horizon size, we 
have also observed that an initial excitation of the internal mode  on the kink  
lasts  for long periods of time, following the same type of behaviour we have seen in Minkowski
space background\footnote{Note that in this case, the bound state profile would be
distorted as well, but in the regime we are interested in, it is sufficient to consider 
the same waveform as in flat space. This shape oscillates in the same way as before
keeping the same physical width over time.}.

In order to show this, we consider an initial configuration made up of a kink in de Sitter 
space, with $H=0.1$, and we add a perturbation in the form of the shape mode with 
amplitude equal to $A(0)= 0.5$. This configuration is evolved for some time in de Sitter space
and then we perform a smooth transition from de Sitter to Minkowski space so that
the final scale factor has grown by a factor of $\sim 30$. 

\begin{figure}[h!]
\includegraphics[width=15.6cm]{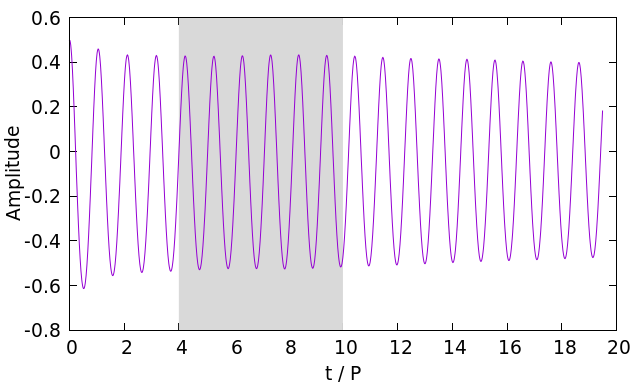}
\caption{Amplitude of the bound  state $A(t)$ over several oscillations evolving 
in an expanding de Sitter background with a small expansion rate $H=0.1$. We then 
transition from de Sitter space to Minkowski space in a smooth way (shaded region) to find that the amplitude
stays pretty much constant. Time is given in units of the dimensionless oscillation period of the shape mode  $P$.}
\label{fig:kink_dS-to-flat}
\end{figure}

There is a subtlety  when performing these transitions between different 
spacetimes. These transitions change by definition the Hubble parameter and
could themselves excite the internal mode of the kink solution. A fast transition 
from a de Sitter universe to flat space, for example, could trigger some
excitation of the internal mode of the soliton. In order to have this effect under 
control, the final transition to Minkowski space has been designed 
to be  a smooth evolution on a time scale larger than the period of the 
oscillation of the bound state. We note that in this way the final result we 
read off from the simulation for the amplitude of the 
internal mode does not depend in any appreciable way on the details
of the transition.

We obtain the amplitude of the shape mode by projecting out on the perturbation, 
as explained in Eq. (\ref{numerical-amplitude})\footnote{In a dynamic spacetime 
we use the same type of projection assuming a fixed physical size of the soliton}. 
As it can be seen in Fig~\ref{fig:kink_dS-to-flat}, the amplitude of the shape mode 
stays quite constant during all this cosmological evolution.

These two numerical experiments allow us to conclude that evolving a kink in 
de Sitter spacetime of Hubble radius a few times larger than its size does not
have much effect on the amplitude of the shape mode.

\subsubsection{Radiation dominated universe}

We have also studied a power law behavior of the scale factor of the form $a(t) \sim t^{1/2}$.
We call this "radiation dominated universe" for its close analogy with the $(3+1)d$ case.
The important point about this type of dynamics is that the Hubble parameter changes
over the course of the evolution in contrast to what happens in de Sitter space. We have 
not been able to find an exact expression for the kink profile in this spacetime. We can 
however approximate the solution by taking an {\it adiabatic} ansatz of the form
\beq
\phi_r(x,t) = \varphi_r \left[ a(t) x \right]\,,
\eeq
which leads to the equation
\beq
(1 - H^2(t) y^2) \varphi_{r}''   =  \varphi_{r}^3 - \varphi_{r} ~.
\eeq

This equation can be used to find an approximate initial configuration for the kink in this
expanding universe by solving it at a particular value of time with
$H=H_0$. This allows us to investigate the possible evolution of the
bound state in this spacetime.

We performed the simulations using our parallel code (so as to ensure a fine enough grid to 
resolve the region of the kink in an expanding background). We also carefully change the 
background from radiation domination to Minkowski, so as to not excite the shape mode in 
the process (as explained earlier).

Our numerical results indicate that starting with a Hubble length somewhat
larger than the width of the kink, the expansion of the universe does not
affect significantly the amplitude of the bound state. This is in agreement with
the results in de Sitter space and the intuition that a small expansion rate
cannot have much influence on the dynamics of the internal mode, whose
frequency is quite higher than $H$.\\

\section{Phase transitions in an expanding background}
\label{phase-transition}

We are  now ready to investigate the main topic of this
paper:  the formation and perdurance of excitations on a kink in a cosmological setting where
the soliton is embedded in a dynamical background.
In this section, we will study numerically  the formation of kinks in a phase transition
and extract the level of excitation of the kinks in this process. 

In order to simulate the phase transition, we will assume that the potential for
the scalar field changes abruptly at some particular moment in time. In other
words, we will consider a time dependent potential that evolves according to the following
prescription \footnote{In this section we re-introduce dimensionful parameters
briefly to clarify the physical content of the theory that we will simulate.}
\begin{equation}
V(\phi) =
\begin{cases} 
\frac{\lambda \eta^4}{4} +  \frac{1}{2} m^2 \phi^2&\textrm{for $t<0$}\\
\frac{\lambda}{4} \left(\phi^2 -\eta^2\right) ^2&\textrm{for $t>0$}
\end{cases}
\end{equation}
where we will take $m^2 = 2\lambda \eta^2$  to be the mass of the field in the first stage of the 
evolution before the phase transition\footnote{Note that
this also corresponds to the mass of the perturbative excitations of the
field around the vacua after the symmetry breaking transition.}.

Furthermore, we will consider the initial conditions of the field to be the ones 
of a thermal state of a massive free field at temperature $T$. There are other ways in which 
 this thermal state can be implemented, and we will discuss some of them in the following section. Here we will
take an approach similar to the one used in \cite{Farhi:2007wj}, where the formation 
of oscillons was discussed. In our lattice field theory representation of the scalar field, this 
means that we will consider the initial state given by

\begin{equation}
\phi\left(x_{j},t=0\right)=\sum_{n=-N/2+1}^{N/2}\frac{1}{\sqrt{2L\omega_{n}}}\left(\alpha_{n}e^{i k_{n}x_{j}}+\alpha_{n}^{*}e^{-ik_{n}x_{j}}\right),
\label{eq:tic phi phys}
\end{equation}

\begin{equation}
\dot \phi\left(x_{j},t=0\right) = \pi\left(x_{j},t=0\right)=\sum_{n=-N/2+1}^{N/2}\frac{1}{i}\sqrt{\frac{\omega_{n}}{2L}}\left(\alpha_{n}e^{i k_{n}x_{j}}-\alpha_{n}^{*}e^{-ik_{n}x_{j}}\right),
\label{eq:tic pi phys}
\end{equation}
\\
where, as usual, we have discretized our box of size $L$ on a lattice with spatial grid size $\Delta x$, so that
the $N$ sites of the lattice, $x_n$, are labeled by the index $n = -N/2+1 ~... ~N/2$. With this notation, the possible wave 
numbers of the reciprocal lattice are given by $k_n = 2\pi n/L$. Given the finite difference scheme described
in Appendix~\ref{numericaldetails}, one can see that the propagating free scalar field modes are parametrized by the frequencies
\begin{equation}
\omega_{n}=\sqrt{\left[\frac{2\sin{\left(\frac{k_{n}\Delta x}{2}\right)}}{\Delta x}\right]^{2}+m^{2}} ~~~.
\label{eq:omega tic}
\end{equation}

Finally, the coefficients of the mode expansion $\alpha_n$ are given by
a Gaussian Random Field whose two point function is given in terms of the thermal spectrum of the
form
\begin{equation}
\langle |\alpha_{n}|^{2}\rangle=\frac{1}{e^{\omega_{n}/T}-1}=\frac{1}{2}\left[\coth{\left(\frac{\omega_{n}}{2T}\right)}-1\right]\,\,.
\label{eq:occupation number tic phys}
\end{equation}

Rewriting this initial configuration in terms of our dimensionless coordinate system, the
spectrum now reads
\begin{equation}
\langle|\tilde{\alpha}_{n}|^{2}\rangle=\frac{1}{2\eta^{2}}\left[\coth{\left(\frac{\tilde{\omega}_{n}}{2\eta^{2}\theta}\right)}-1\right]\,\,,
\label{eq:occupation number tic}
\end{equation}
where, as before, the quantities with tildes are dimensionless so  $\tilde{\omega} = \omega / (\sqrt{\lambda} \eta)$ and the pre-factor $\eta^2$
is introduced to ensure that our expansion of the field is in agreement with our coordinates. This thermal state is therefore fully specified by two dimensionless parameters, namely,
\beq
\theta = \frac{T}{\sqrt{\lambda} \eta^3}   ~~~~~~~~\text{and}~~ ~~~~~~~ \eta ~.
\eeq

The origin of the first one is clear since it controls the ratio of the typical thermal energy to the energy scale
associated with the kink solution (see for example the expression of the mass of the soliton in Eq. (\ref{mass-kink})).
The appearance of $\eta$ explicitly can be traced to the fact that our distribution takes into account the
quantum effects of the Bose-Einstein distribution. Indeed, in the classical limit where $\omega_n << T$ 
the distribution becomes the Rayleigh-Jeans one and the explicit dependence on $\eta$ disappears.

In the following simulations we will choose the parameters so that the amount of initial energy in the thermal
state is subleading with respect to the potential energy of the vacuum state. Moreover, we will also
take the quantum cutoff of the Bose-Einstein distribution at high frequencies such that this corresponds
to a wavelength slightly larger than the lattice spacing. Both these facts will ensure that our
initial conditions are dominated by a classical regime so that the use of classical 
equations of motion to obtain the distribution of defects in the transition is justified.
We will fix $\theta$ and $\eta$ so that both these requirements are satisfied in our simulations.

\subsection{Evolving in an expanding background}

The formation of domain wall solitons in a cosmological phase transition involves, of course,
an expanding universe. In that regard it is interesting to perform our
numerical simulation in this type of background. However, there is  another reason
 to do this simulation in an expanding universe: it provides the simulation with a natural friction term.
 In fact,
 the background energy density present at the beginning of the simulation is
very large compared to the average energy density expected
as a final product of the transition in order to identify the individual kinks. This means 
that a mechanism for dissipation of this extra energy is needed. Without 
an efficient dissipation, kinks would acquire large kinetic energies and would 
annihilate with each other easily, leaving behind a large collection of perturbative 
excitations. Even if we manage to create a simulation with large enough volume where
some of the kinks would survive for long time, the background energy would
continuously excite the internal modes of the kink. In such case it would not be 
clear when to stop the simulation to obtain an accurate evaluation of the 
level of excitation of any internal modes present in the  defects.

One can imagine fixing this problem by simulating the transition
with some added friction term, but this is somewhat arbitrary since one would
be able to control the final result by adjusting the amplitude and duration of the
friction. This is why  performing the simulations of the transition in an expanding 
spacetime is advantageous. With the natural friction it provides,  the background energy density
slowly depletes. Furthermore, as we
showed in our previous section, taking a small enough expansion rate does not affect much  the
amplitude of the internal mode of the kink solution.

This cosmological friction term  also induces a horizon size, so kinks and 
anti-kinks present at distances larger than this distance are not
to annihilate. Their velocities are redshifted until they are comoving with
the background. 

All these effects will allow us to obtain a much more realistic final configuration
of the transition where kinks are well separated and almost at rest with respect to the
simulation grid. This is a simple configuration that can be used to extract 
the amplitude of the internal mode of the kinks.

As a final step in the simulation, we also consider a smooth transition of the
background to $1+1$ dimensional flat Minkowski spacetime, as explained earlier. We  
evolve the simulation over large periods of time with absorbing boundary
conditions, tracking the amplitude of the shape modes of the kinks.

\subsection{Results}

We have run several simulations of $N=5000$ points with $\Delta x = 0.01$ with an initial thermal 
state that corresponds to a dimensionless temperature of $\theta = 10^{-3}$
and a value of the parameter $\eta = 250$. For these values of the parameters, the initial amount of thermal
energy is of the order of $15\%$ of the total background energy density. This also means that
the thermal modes are substantially suppressed for wavelengths smaller that  $\lambda_T \approx 10 \Delta x$. 
So the thermal spectrum at the scale of the soliton is correctly represented by a classic
thermal state.

\begin{figure}[h!]
\begingroup
  \makeatletter
  \providecommand\color[2][]{%
    \GenericError{(gnuplot) \space\space\space\@spaces}{%
      Package color not loaded in conjunction with
      terminal option `colourtext'%
    }{See the gnuplot documentation for explanation.%
    }{Either use 'blacktext' in gnuplot or load the package
      color.sty in LaTeX.}%
    \renewcommand\color[2][]{}%
  }%
  \providecommand\includegraphics[2][]{%
    \GenericError{(gnuplot) \space\space\space\@spaces}{%
      Package graphicx or graphics not loaded%
    }{See the gnuplot documentation for explanation.%
    }{The gnuplot epslatex terminal needs graphicx.sty or graphics.sty.}%
    \renewcommand\includegraphics[2][]{}%
  }%
  \providecommand\rotatebox[2]{#2}%
  \@ifundefined{ifGPcolor}{%
    \newif\ifGPcolor
    \GPcolorfalse
  }{}%
  \@ifundefined{ifGPblacktext}{%
    \newif\ifGPblacktext
    \GPblacktexttrue
  }{}%
  \let\gplgaddtomacro\g@addto@macro
  \gdef\gplbacktext{}%
  \gdef\gplfronttext{}%
  \makeatother
  \ifGPblacktext
    \def\colorrgb#1{}%
    \def\colorgray#1{}%
  \else
    \ifGPcolor
      \def\colorrgb#1{\color[rgb]{#1}}%
      \def\colorgray#1{\color[gray]{#1}}%
      \expandafter\def\csname LTw\endcsname{\color{white}}%
      \expandafter\def\csname LTb\endcsname{\color{black}}%
      \expandafter\def\csname LTa\endcsname{\color{black}}%
      \expandafter\def\csname LT0\endcsname{\color[rgb]{1,0,0}}%
      \expandafter\def\csname LT1\endcsname{\color[rgb]{0,1,0}}%
      \expandafter\def\csname LT2\endcsname{\color[rgb]{0,0,1}}%
      \expandafter\def\csname LT3\endcsname{\color[rgb]{1,0,1}}%
      \expandafter\def\csname LT4\endcsname{\color[rgb]{0,1,1}}%
      \expandafter\def\csname LT5\endcsname{\color[rgb]{1,1,0}}%
      \expandafter\def\csname LT6\endcsname{\color[rgb]{0,0,0}}%
      \expandafter\def\csname LT7\endcsname{\color[rgb]{1,0.3,0}}%
      \expandafter\def\csname LT8\endcsname{\color[rgb]{0.5,0.5,0.5}}%
    \else
      \def\colorrgb#1{\color{black}}%
      \def\colorgray#1{\color[gray]{#1}}%
      \expandafter\def\csname LTw\endcsname{\color{white}}%
      \expandafter\def\csname LTb\endcsname{\color{black}}%
      \expandafter\def\csname LTa\endcsname{\color{black}}%
      \expandafter\def\csname LT0\endcsname{\color{black}}%
      \expandafter\def\csname LT1\endcsname{\color{black}}%
      \expandafter\def\csname LT2\endcsname{\color{black}}%
      \expandafter\def\csname LT3\endcsname{\color{black}}%
      \expandafter\def\csname LT4\endcsname{\color{black}}%
      \expandafter\def\csname LT5\endcsname{\color{black}}%
      \expandafter\def\csname LT6\endcsname{\color{black}}%
      \expandafter\def\csname LT7\endcsname{\color{black}}%
      \expandafter\def\csname LT8\endcsname{\color{black}}%
    \fi
  \fi
    \setlength{\unitlength}{0.0500bp}%
    \ifx\gptboxheight\undefined%
      \newlength{\gptboxheight}%
      \newlength{\gptboxwidth}%
      \newsavebox{\gptboxtext}%
    \fi%
    \setlength{\fboxrule}{0.5pt}%
    \setlength{\fboxsep}{1pt}%
\begin{picture}(8844.00,9636.00)%
    \gplgaddtomacro\gplbacktext{%
      \csname LTb\endcsname
      \put(1078,7128){\makebox(0,0)[r]{\strut{}$-0.06$}}%
      \put(1078,7423){\makebox(0,0)[r]{\strut{}$-0.04$}}%
      \put(1078,7719){\makebox(0,0)[r]{\strut{}$-0.02$}}%
      \put(1078,8014){\makebox(0,0)[r]{\strut{}$0$}}%
      \put(1078,8309){\makebox(0,0)[r]{\strut{}$0.02$}}%
      \put(1078,8604){\makebox(0,0)[r]{\strut{}$0.04$}}%
      \put(1078,8900){\makebox(0,0)[r]{\strut{}$0.06$}}%
      \put(1078,9195){\makebox(0,0)[r]{\strut{}$0.08$}}%
      \put(1934,6908){\makebox(0,0){\strut{}$-20$}}%
      \put(3381,6908){\makebox(0,0){\strut{}$-10$}}%
      \put(4829,6908){\makebox(0,0){\strut{}$0$}}%
      \put(6276,6908){\makebox(0,0){\strut{}$10$}}%
      \put(7723,6908){\makebox(0,0){\strut{}$20$}}%
    }%
    \gplgaddtomacro\gplfronttext{%
      \csname LTb\endcsname
      \put(198,8161){\rotatebox{-270}{\makebox(0,0){\Large $\phi$}}}%
      \put(4828,6578){\makebox(0,0){\Large x}}%
    }%
    \gplgaddtomacro\gplbacktext{%
      \csname LTb\endcsname
      \put(946,3916){\makebox(0,0)[r]{\strut{}$-1.5$}}%
      \put(946,4329){\makebox(0,0)[r]{\strut{}$-1$}}%
      \put(946,4743){\makebox(0,0)[r]{\strut{}$-0.5$}}%
      \put(946,5156){\makebox(0,0)[r]{\strut{}$0$}}%
      \put(946,5570){\makebox(0,0)[r]{\strut{}$0.5$}}%
      \put(946,5983){\makebox(0,0)[r]{\strut{}$1$}}%
      \put(1815,3696){\makebox(0,0){\strut{}$-20$}}%
      \put(3289,3696){\makebox(0,0){\strut{}$-10$}}%
      \put(4763,3696){\makebox(0,0){\strut{}$0$}}%
      \put(6236,3696){\makebox(0,0){\strut{}$10$}}%
      \put(7710,3696){\makebox(0,0){\strut{}$20$}}%
    }%
    \gplgaddtomacro\gplfronttext{%
      \csname LTb\endcsname
      \put(198,4949){\rotatebox{-270}{\makebox(0,0){\Large $\phi$}}}%
      \put(4762,3366){\makebox(0,0){\Large x}}%
    }%
    \gplgaddtomacro\gplbacktext{%
      \csname LTb\endcsname
      \put(946,704){\makebox(0,0)[r]{\strut{}$-1.5$}}%
      \put(946,1049){\makebox(0,0)[r]{\strut{}$-1$}}%
      \put(946,1393){\makebox(0,0)[r]{\strut{}$-0.5$}}%
      \put(946,1738){\makebox(0,0)[r]{\strut{}$0$}}%
      \put(946,2083){\makebox(0,0)[r]{\strut{}$0.5$}}%
      \put(946,2427){\makebox(0,0)[r]{\strut{}$1$}}%
      \put(946,2772){\makebox(0,0)[r]{\strut{}$1.5$}}%
      \put(1815,484){\makebox(0,0){\strut{}$-20$}}%
      \put(3289,484){\makebox(0,0){\strut{}$-10$}}%
      \put(4763,484){\makebox(0,0){\strut{}$0$}}%
      \put(6236,484){\makebox(0,0){\strut{}$10$}}%
      \put(7710,484){\makebox(0,0){\strut{}$20$}}%
    }%
    \gplgaddtomacro\gplfronttext{%
      \csname LTb\endcsname
      \put(198,1738){\rotatebox{-270}{\makebox(0,0){\Large $\phi$}}}%
      \put(4762,154){\makebox(0,0){\Large x}}%
    }%
    \gplbacktext
    \put(0,0){\includegraphics{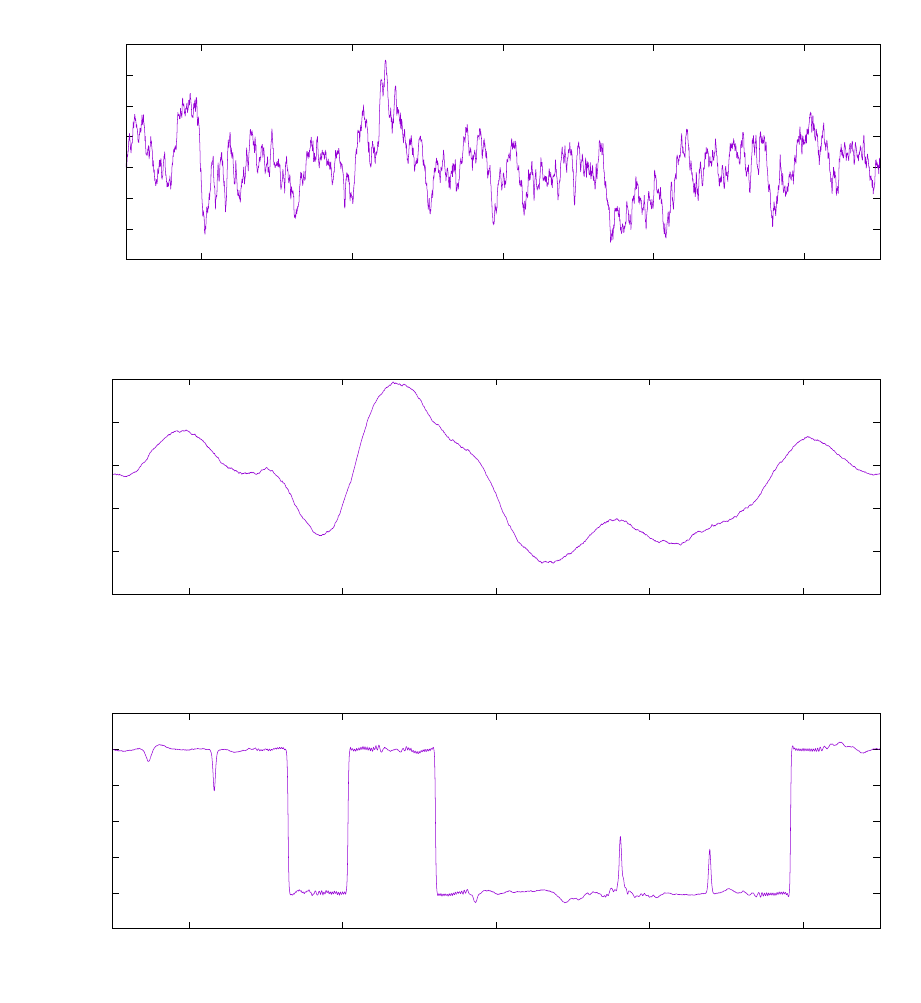}}%
    \gplfronttext
  \end{picture}%
\endgroup
\caption{Three snapshots of the field profile during the phase transition. The top one corresponds to the
initial conditions. The middle one is at some intermediate step during the de Sitter phase and the bottom 
one is at cosmic time $t=140$ (27.3 in units of the oscillation period of the shape 
mode $P$). This last instant of time already corresponds to the Minkowski stage. Several kinks and 
antikinks have been created.}
\label{fig:field pt}
\end{figure}

We have evolved these initial configurations in an expanding de Sitter background whose
Hubble constant in dimensionless units was set to $H_i = 0.04$. This means that the associated horizon
size is substantially larger than the size of the kink soliton. Following the numerical
experiments we described in the previous section, we are confident that any initial
amplitude on the internal mode of the kink produced by the dynamics of the phase transition 
will survive in this background. Moreover, the background energy density of perturbative 
excitations is heavily suppressed during this time.

We show in Fig.~\ref{fig:field pt}  a few snapshots of the field at different times for a particular realization
of this setup. We notice that as time evolves
the field settles on one of the vacua in different regions of space. This leads to the formation
of a "network" of alternating kinks and anti-kinks. Furthermore, this transition also leads to the
generation of localized oscillating lumps around each of the vacua. These are nothing more than the 
breather type solutions that we discussed in previous sections. They are readily produced
by the relaxation mechanism of the system directly and also by the collision and merging
of nearby kinks and anti-kinks.

We let the system evolve until $t\approx 80$, where we relax the expansion of the universe
smoothly into flat Minkowski space similarly to what we did in the previous section. By this time
the background energy has already decreased substantially and one can observe the kinks
forming. We also use absorbing boundary conditions so that radiation is allowed to be absorbed 
by the boundary and leave the simulation. The late time result of this simulation is a system 
of well separated kinks and anti-kinks, pretty much at rest with respect to one another, plus a 
collection of breathers. 

\begin{figure}[h!]
\includegraphics[width=\textwidth]{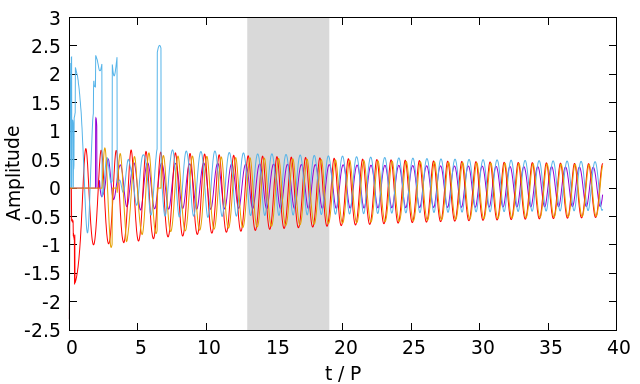}
\caption{Amplitude of the shape mode $A(t)$ as a function of time, for each of the four kinks in 
Fig.~\ref{fig:field pt}. The color correspondence is purple, red, blue and orange as we encounter 
the kinks going from left to right in Fig.~\ref{fig:field pt}. Time is given in units of the oscillation 
period of the shape mode $P$, and the shaded region represents the smooth transition from 
de Sitter to Minkowski space.}
\label{fig:asm pt}
\end{figure}

We concentrate on each of the kinks and investigate their level of excitation by extracting the amplitude of the internal mode for each kink using the expression given
by Eq. (\ref{numerical-amplitude}) locally. The results for the amplitudes of the kinks in 
one single realization are displayed in Fig.~\ref{fig:asm pt}. We notice that the resulting amplitudes of the 
internal modes for the different kinks are quite similar.

In order to find the average value of this amplitude of the shape mode bound state, we have run $500$ 
realizations with a thermal distribution of initial perturbations with $\theta=10^{-3}$ to get that
\beq
\langle ~\hat A~\rangle_{\text{formation}} = 0.5 \pm 0.1~.
\eeq

We have also investigated the dependence of this quantity on the initial conditions by changing the
temperature of the initial thermal state so that the percentage of extra initial energy due to this
perturbations was smaller by a factor of $2$. However, the final result for the average amplitude of the
bound state of the kink was pretty insensitive to these changes. This may be attributed to the 
fact that the relevant energy scale in the process of the kink formation is the initial background
energy associated with the initial vacuum state. In other words, the initial thermal fluctuations
are important to induce inhomogeneities that lead to the formation of kinks and anti-kinks but
not to the final energy stored in them.

We have also run simulations where the expanding spacetime was described by a "radiation
domination" scale factor. The results are qualitatively similar to the case of de Sitter space.

The movies produced from these simulations
show that the excitation of the kinks is due to
mainly two effects. The first one is the interaction of the kinks with the perturbative excitations
on both sides of the soliton that pass through them. At a linear level, these modes should not
excite the kink.  However, at a non-linear level, these waves produce a non-zero excitation of
the zero mode as well as the shape mode. We have actually tested this by recreating this type
of interactions on isolated kinks that we irradiated with wave-packets of different frequencies.
These scattering experiments indeed produced some motion of the kinks as well as an
excitation of the internal mode.

The second mechanism for kink excitation is more non-linear in nature and has to do with the
collision between breathers and kink-solitons. As we described in Section~\ref{the-breather}, one can consider
the internal shape mode as a breather trapped on a kink. In fact, one can devise a low amplitude 
breather with very similar characteristics as the internal mode. Therefore, it should not come as
a surprise   that the collisions of breathers with kinks are a good place to see the
amplification of the energy stored in the internal mode.

\begin{figure}[h!]
\begingroup
  \makeatletter
  \providecommand\color[2][]{%
    \GenericError{(gnuplot) \space\space\space\@spaces}{%
      Package color not loaded in conjunction with
      terminal option `colourtext'%
    }{See the gnuplot documentation for explanation.%
    }{Either use 'blacktext' in gnuplot or load the package
      color.sty in LaTeX.}%
    \renewcommand\color[2][]{}%
  }%
  \providecommand\includegraphics[2][]{%
    \GenericError{(gnuplot) \space\space\space\@spaces}{%
      Package graphicx or graphics not loaded%
    }{See the gnuplot documentation for explanation.%
    }{The gnuplot epslatex terminal needs graphicx.sty or graphics.sty.}%
    \renewcommand\includegraphics[2][]{}%
  }%
  \providecommand\rotatebox[2]{#2}%
  \@ifundefined{ifGPcolor}{%
    \newif\ifGPcolor
    \GPcolorfalse
  }{}%
  \@ifundefined{ifGPblacktext}{%
    \newif\ifGPblacktext
    \GPblacktexttrue
  }{}%
  \let\gplgaddtomacro\g@addto@macro
  \gdef\gplbacktext{}%
  \gdef\gplfronttext{}%
  \makeatother
  \ifGPblacktext
    \def\colorrgb#1{}%
    \def\colorgray#1{}%
  \else
    \ifGPcolor
      \def\colorrgb#1{\color[rgb]{#1}}%
      \def\colorgray#1{\color[gray]{#1}}%
      \expandafter\def\csname LTw\endcsname{\color{white}}%
      \expandafter\def\csname LTb\endcsname{\color{black}}%
      \expandafter\def\csname LTa\endcsname{\color{black}}%
      \expandafter\def\csname LT0\endcsname{\color[rgb]{1,0,0}}%
      \expandafter\def\csname LT1\endcsname{\color[rgb]{0,1,0}}%
      \expandafter\def\csname LT2\endcsname{\color[rgb]{0,0,1}}%
      \expandafter\def\csname LT3\endcsname{\color[rgb]{1,0,1}}%
      \expandafter\def\csname LT4\endcsname{\color[rgb]{0,1,1}}%
      \expandafter\def\csname LT5\endcsname{\color[rgb]{1,1,0}}%
      \expandafter\def\csname LT6\endcsname{\color[rgb]{0,0,0}}%
      \expandafter\def\csname LT7\endcsname{\color[rgb]{1,0.3,0}}%
      \expandafter\def\csname LT8\endcsname{\color[rgb]{0.5,0.5,0.5}}%
    \else
      \def\colorrgb#1{\color{black}}%
      \def\colorgray#1{\color[gray]{#1}}%
      \expandafter\def\csname LTw\endcsname{\color{white}}%
      \expandafter\def\csname LTb\endcsname{\color{black}}%
      \expandafter\def\csname LTa\endcsname{\color{black}}%
      \expandafter\def\csname LT0\endcsname{\color{black}}%
      \expandafter\def\csname LT1\endcsname{\color{black}}%
      \expandafter\def\csname LT2\endcsname{\color{black}}%
      \expandafter\def\csname LT3\endcsname{\color{black}}%
      \expandafter\def\csname LT4\endcsname{\color{black}}%
      \expandafter\def\csname LT5\endcsname{\color{black}}%
      \expandafter\def\csname LT6\endcsname{\color{black}}%
      \expandafter\def\csname LT7\endcsname{\color{black}}%
      \expandafter\def\csname LT8\endcsname{\color{black}}%
    \fi
  \fi
    \setlength{\unitlength}{0.0500bp}%
    \ifx\gptboxheight\undefined%
      \newlength{\gptboxheight}%
      \newlength{\gptboxwidth}%
      \newsavebox{\gptboxtext}%
    \fi%
    \setlength{\fboxrule}{0.5pt}%
    \setlength{\fboxsep}{1pt}%
\begin{picture}(9060.00,5660.00)%
    \gplgaddtomacro\gplbacktext{%
      \csname LTb\endcsname
      \put(747,595){\makebox(0,0)[r]{\strut{}$-1.5$}}%
      \csname LTb\endcsname
      \put(747,1292){\makebox(0,0)[r]{\strut{}$-1$}}%
      \csname LTb\endcsname
      \put(747,1989){\makebox(0,0)[r]{\strut{}$-0.5$}}%
      \csname LTb\endcsname
      \put(747,2686){\makebox(0,0)[r]{\strut{}$0$}}%
      \csname LTb\endcsname
      \put(747,3382){\makebox(0,0)[r]{\strut{}$0.5$}}%
      \csname LTb\endcsname
      \put(747,4079){\makebox(0,0)[r]{\strut{}$1$}}%
      \csname LTb\endcsname
      \put(747,4776){\makebox(0,0)[r]{\strut{}$1.5$}}%
      \csname LTb\endcsname
      \put(747,5473){\makebox(0,0)[r]{\strut{}$2$}}%
      \csname LTb\endcsname
      \put(1639,409){\makebox(0,0){\strut{}$-20$}}%
      \csname LTb\endcsname
      \put(3220,409){\makebox(0,0){\strut{}$-10$}}%
      \csname LTb\endcsname
      \put(4801,409){\makebox(0,0){\strut{}$0$}}%
      \csname LTb\endcsname
      \put(6382,409){\makebox(0,0){\strut{}$10$}}%
      \csname LTb\endcsname
      \put(7963,409){\makebox(0,0){\strut{}$20$}}%
    }%
    \gplgaddtomacro\gplfronttext{%
      \csname LTb\endcsname
      \put(153,3034){\rotatebox{-270}{\makebox(0,0){\Large $\phi$}}}%
      \csname LTb\endcsname
      \put(4801,130){\makebox(0,0){\Large x}}%
    }%
    \gplbacktext
    \put(0,0){\includegraphics{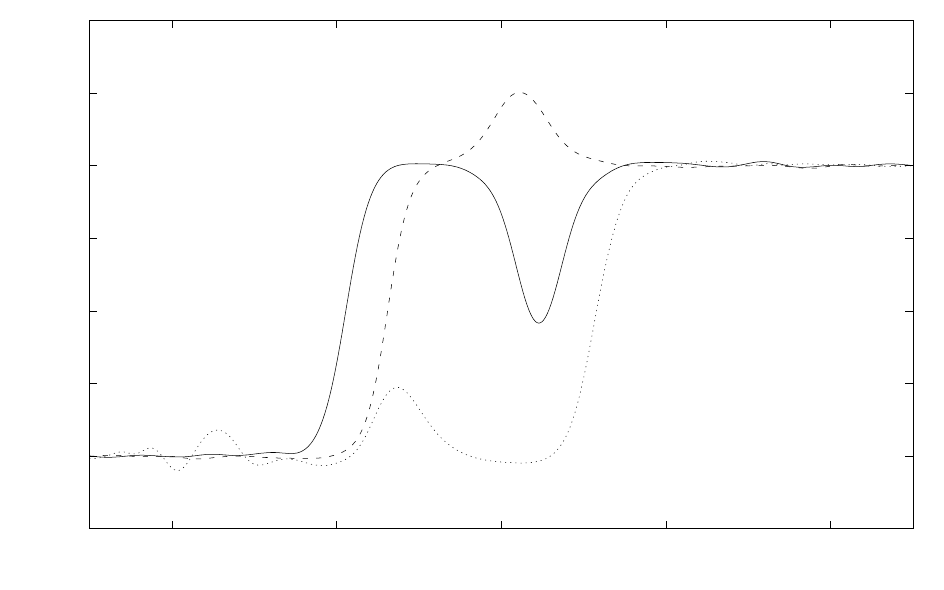}}%
    \gplfronttext
  \end{picture}%
\endgroup
\caption{Thee different snapshots of the field showing the interaction between a kink and a breather.
The solid line is the earlier configuration, the dashed line illustrates the two objects coming together, and the
dotted line represents the result after they have passed through one another. The resulting amplitude
of the shape mode in this scattering process is given in Fig.~\ref{fig:asm_oscillon}.}
\label{fig:field-kink--oscillon}
\end{figure}

\begin{figure}[h!]
\includegraphics[scale=0.7]{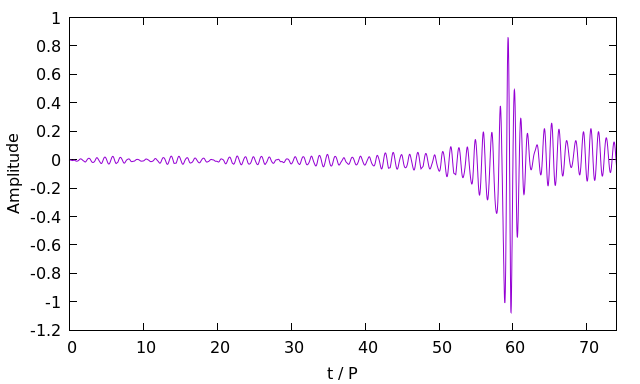}
\caption{Amplitude of the shape mode $A(t)$ as a function of time, during a collision of a kink with a 
breather. Time is given in units of the oscillation period of the shape mode $P$.}
\label{fig:asm_oscillon}
\end{figure}

We show in Fig.~\ref{fig:field-kink--oscillon}  three different snapshots of the field configurations for one such type 
of event. The initial configuration shows a kink and a breather at a short distance from one another. 
As times passes, the breather becomes closer to the kink and overlaps with it for a while. One can see that this
triggers a big resonance effect on the amplitude of the internal mode, so the final state
of the kink is substantially more energetic. See Fig.~\ref{fig:asm_oscillon} for the evolution of the amplitude
of the internal mode as a function of time \footnote{The kink translational mode is also excited
in this process.}.

The combination of these two effects leads to the final amplitude of the shape mode on a typical
phase transition.  The kinks seem to have at formation around $20\%$ more
energy than the lowest energy configuration. This is not a negligible amount of energy, and it 
could easily have important consequences for the subsequent evolution. In particular, as we 
showed in Section~\ref{lifetime}, this kind of extra energy will stay in the kink for long periods 
of time compared to the natural time scale of the kink, which is given by $\tau \sim 1/m$. 

\section{Heating up the kink }
\label{thermal-bath}

Another interesting scenario  is the possible excitation of the 
solitons due to their interaction with a thermal bath. This could happen, for example,
in the formation of defects in a cosmological setting during reheating \cite{Tkachev:1995md}. Alternatively,
this could also be useful to estimate the possible degree of excitation in numerical
simulations where there is a substantial amount of radiation in the background.\footnote{In this
paper we have run the simulations with a thermal spectrum but could easily extend these
ideas to include some other spectra.}

There are several ways in which a thermal background can be simulated.
Following the description in the previous section, we can approximate the solution of 
a kink in a thermal bath as a sum of the lowest energy configuration for the soliton together with
a state of the form given by Eq. (\ref{eq:tic phi phys}).

One could object that the presence of a kink will distort the spectrum of the
perturbations around the solution, possibly affecting the amplitude of the
bound state that we want to find. We have addressed this concern by
simulating a thermal interaction of the kink with a bath using the Metropolis
algorithm \cite{Metropolis}. We give the details of this procedure in Appendix~\ref{Metropolis}.
This is a much more expensive procedure computationally than the one described above,  since 
 the lattice needs to be swept many times over before a
configuration that resembles the thermal state is reached\footnote{This method has been previously
used in the literature in similar models like \cite{Grigoriev:1988bd}.}. We have done this
for several different realizations and compared the results with the
analytic setup of the thermal state described in the previous section
to find that both these methods lead to comparable results for the
amplitude of the shape mode.

\begin{figure}[h!]
\begingroup
  \makeatletter
  \providecommand\color[2][]{%
    \GenericError{(gnuplot) \space\space\space\@spaces}{%
      Package color not loaded in conjunction with
      terminal option `colourtext'%
    }{See the gnuplot documentation for explanation.%
    }{Either use 'blacktext' in gnuplot or load the package
      color.sty in LaTeX.}%
    \renewcommand\color[2][]{}%
  }%
  \providecommand\includegraphics[2][]{%
    \GenericError{(gnuplot) \space\space\space\@spaces}{%
      Package graphicx or graphics not loaded%
    }{See the gnuplot documentation for explanation.%
    }{The gnuplot epslatex terminal needs graphicx.sty or graphics.sty.}%
    \renewcommand\includegraphics[2][]{}%
  }%
  \providecommand\rotatebox[2]{#2}%
  \@ifundefined{ifGPcolor}{%
    \newif\ifGPcolor
    \GPcolorfalse
  }{}%
  \@ifundefined{ifGPblacktext}{%
    \newif\ifGPblacktext
    \GPblacktexttrue
  }{}%
  \let\gplgaddtomacro\g@addto@macro
  \gdef\gplbacktext{}%
  \gdef\gplfronttext{}%
  \makeatother
  \ifGPblacktext
    \def\colorrgb#1{}%
    \def\colorgray#1{}%
  \else
    \ifGPcolor
      \def\colorrgb#1{\color[rgb]{#1}}%
      \def\colorgray#1{\color[gray]{#1}}%
      \expandafter\def\csname LTw\endcsname{\color{white}}%
      \expandafter\def\csname LTb\endcsname{\color{black}}%
      \expandafter\def\csname LTa\endcsname{\color{black}}%
      \expandafter\def\csname LT0\endcsname{\color[rgb]{1,0,0}}%
      \expandafter\def\csname LT1\endcsname{\color[rgb]{0,1,0}}%
      \expandafter\def\csname LT2\endcsname{\color[rgb]{0,0,1}}%
      \expandafter\def\csname LT3\endcsname{\color[rgb]{1,0,1}}%
      \expandafter\def\csname LT4\endcsname{\color[rgb]{0,1,1}}%
      \expandafter\def\csname LT5\endcsname{\color[rgb]{1,1,0}}%
      \expandafter\def\csname LT6\endcsname{\color[rgb]{0,0,0}}%
      \expandafter\def\csname LT7\endcsname{\color[rgb]{1,0.3,0}}%
      \expandafter\def\csname LT8\endcsname{\color[rgb]{0.5,0.5,0.5}}%
    \else
      \def\colorrgb#1{\color{black}}%
      \def\colorgray#1{\color[gray]{#1}}%
      \expandafter\def\csname LTw\endcsname{\color{white}}%
      \expandafter\def\csname LTb\endcsname{\color{black}}%
      \expandafter\def\csname LTa\endcsname{\color{black}}%
      \expandafter\def\csname LT0\endcsname{\color{black}}%
      \expandafter\def\csname LT1\endcsname{\color{black}}%
      \expandafter\def\csname LT2\endcsname{\color{black}}%
      \expandafter\def\csname LT3\endcsname{\color{black}}%
      \expandafter\def\csname LT4\endcsname{\color{black}}%
      \expandafter\def\csname LT5\endcsname{\color{black}}%
      \expandafter\def\csname LT6\endcsname{\color{black}}%
      \expandafter\def\csname LT7\endcsname{\color{black}}%
      \expandafter\def\csname LT8\endcsname{\color{black}}%
    \fi
  \fi
    \setlength{\unitlength}{0.0500bp}%
    \ifx\gptboxheight\undefined%
      \newlength{\gptboxheight}%
      \newlength{\gptboxwidth}%
      \newsavebox{\gptboxtext}%
    \fi%
    \setlength{\fboxrule}{0.5pt}%
    \setlength{\fboxsep}{1pt}%
\begin{picture}(8844.00,5668.00)%
    \gplgaddtomacro\gplbacktext{%
      \csname LTb\endcsname
      \put(946,704){\makebox(0,0)[r]{\strut{}$-1.5$}}%
      \put(946,1495){\makebox(0,0)[r]{\strut{}$-1$}}%
      \put(946,2285){\makebox(0,0)[r]{\strut{}$-0.5$}}%
      \put(946,3076){\makebox(0,0)[r]{\strut{}$0$}}%
      \put(946,3866){\makebox(0,0)[r]{\strut{}$0.5$}}%
      \put(946,4657){\makebox(0,0)[r]{\strut{}$1$}}%
      \put(946,5447){\makebox(0,0)[r]{\strut{}$1.5$}}%
      \put(1078,484){\makebox(0,0){\strut{}$-25$}}%
      \put(1815,484){\makebox(0,0){\strut{}$-20$}}%
      \put(2552,484){\makebox(0,0){\strut{}$-15$}}%
      \put(3289,484){\makebox(0,0){\strut{}$-10$}}%
      \put(4026,484){\makebox(0,0){\strut{}$-5$}}%
      \put(4763,484){\makebox(0,0){\strut{}$0$}}%
      \put(5499,484){\makebox(0,0){\strut{}$5$}}%
      \put(6236,484){\makebox(0,0){\strut{}$10$}}%
      \put(6973,484){\makebox(0,0){\strut{}$15$}}%
      \put(7710,484){\makebox(0,0){\strut{}$20$}}%
      \put(8447,484){\makebox(0,0){\strut{}$25$}}%
    }%
    \gplgaddtomacro\gplfronttext{%
      \csname LTb\endcsname
      \put(198,3075){\rotatebox{-270}{\makebox(0,0){\Large $\phi$}}}%
      \put(4762,154){\makebox(0,0){\Large x}}%
    }%
    \gplbacktext
    \put(0,0){\includegraphics{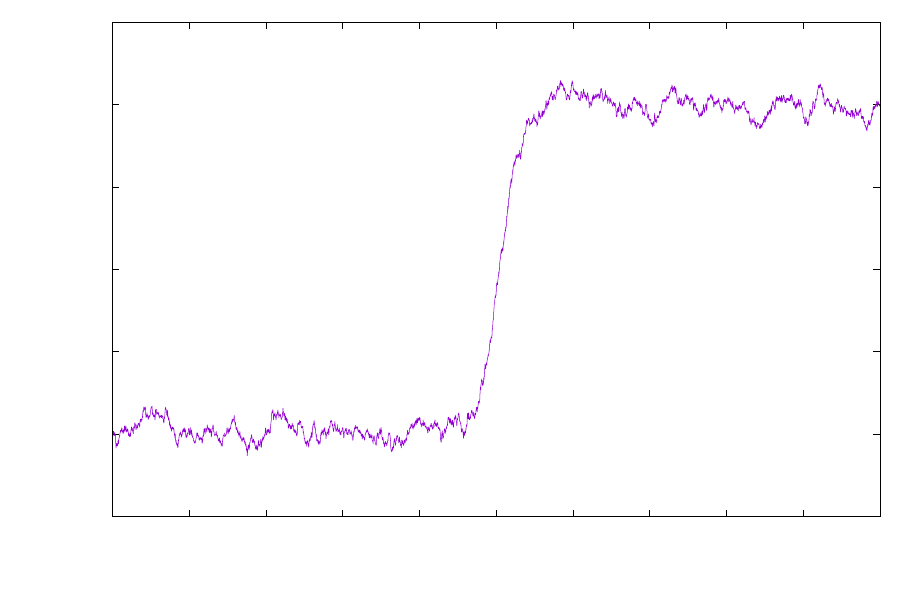}}%
    \gplfronttext
  \end{picture}%
\endgroup
\caption{Initial field configuration. It has been generated by means of the Metropolis 
algorithm, at temperature $\theta=0.01$.}
\label{fig:thermal kink}
\end{figure}

We show in Fig.~\ref{fig:thermal kink}  a typical example of such initial configuration of a kink
at a dimensionless temperature $\theta = 0.01$. Another important point to note is that
the interaction with the thermal fluctuations imprints an initial velocity for the kink as well. 
We have checked that the distribution of initial velocities are in agreement with the
expectation of a particle of mass $M_k$ in a thermal state of temperature $T$.
In practice, this initial velocity complicates the simulation in Minkowski space since for
large temperatures the kink will eventually reach the boundary of our box and leave it.

\begin{figure}[h!]
\includegraphics[width=15.6cm]{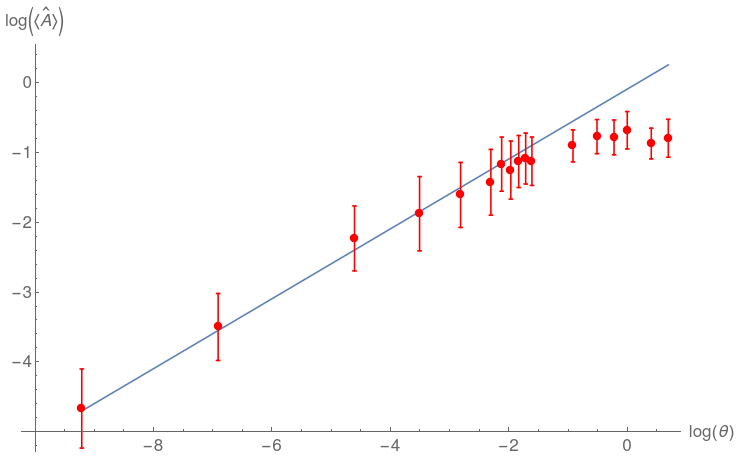}
\caption{Log-Log plot for the average amplitude of the shape mode $\langle \hat A(t) \rangle$, 
as a function of temperature. The amplitude is observed to tend to a constant value 
$\langle \hat A \rangle\approx0.5$ in the high temperature limit.}
\label{fig:amplitud-versus-T}
\end{figure}

We have therefore decided to simulate the initial thermal state in a de Sitter
background with a small Hubble expansion rate. As we learned before, this
does not have much of an effect on the amplitude of the shape mode, and 
 it  slows down the kink so that it can be  simulated for large periods
of time. 
This process allows us to reliably extract the value of the 
amplitude of the internal bound state. We have run
$100$ realizations for each temperature ranging from $ 10^{-4} <  \theta < 2$ 
and obtained the average value of the amplitude of the kink. We have
plotted the results in Fig.~\ref{fig:amplitud-versus-T}. 

The analytic estimate of the average 
value of the amplitude as a function of the dimensionless temperature is described in  Appendix~\ref{analitic-projection}.
This is done by looking at the projection of the thermal fluctuations over the 
bound state mode. This simple description implies that  
$\langle \hat A \rangle \propto \theta^{1/2}$. This is a  good 
fit to our results at low temperature but deviates
from the actual numerical results at higher values of the temperature.
This is a somewhat expected result since the argument of the
analytic estimate is based on purely linear approximation of the
background radiation. However, as one approaches large temperatures, 
the amplitude of the thermal perturbations is not so small.

As we increase the temperature, we notice that the energy stored
in the thermal fluctuations is not negligible compared to the energy
of the kink. This means that in some realizations there are fluctuations
that generate a kink-antikink pair directly from the background. For
temperatures above $\theta = 0.2$, the number of these pairs
is large.

We have performed these numerical simulations for temperatures
as large as $\theta = 2$. In this case, the initial kink is irrelevant 
since the energy is so high that the typical fluctuations can 
go over the maximum of the potential. However, simulating this
in an expanding background and with absorbing boundary
conditions, we are able to end up with a collection of isolated kinks.
We see
that the amplitude of the kinks formed this way is of the order of $\langle \hat A \rangle \sim 0.5$.
This is quite similar to the value we obtained in the previous
simulations of the phase transitions.

It is interesting to note that one does not seem to be able to reach
the strong non-linear regime for the shape mode amplitude in our
numerical simulations. This is indeed the case for both models, the phase
transition and the thermal bath interaction.  One is tempted to speculate
that even though the bound state could store much more
energy, in practice, natural initial conditions for cosmology
do not lead to such situations.

One can also simulate this interaction of the kink and a thermal
bath using the Langevin equations (see for example \cite{Alford:1991qg}).
It would be interesting to see if the results for the amplitude
of the bound state in this case are in agreement with our
simulations. We leave this question for future investigations.

\section{Conclusions}

One of the most interesting properties of solitonic solutions for cosmological applications 
is their long lifetimes compared to the fundamental time scale of the underlying theory. This is typically
due to the fact that they correspond to the lowest energy configuration associated
with a charge, either topological or non-topological,  that ensures their
stability.

In many field theory models, this long perdurance is also shared by other type of localized configurations
that go by different names, such as breathers or oscillons. The reason for these other
objects to have such a long lifetime is different. They are oscillating field configurations
whose frequencies are below the frequencies of the propagating modes
in the vacuum.

Here we study a sort of hybrid configuration, one that owes its
long duration to dynamical reasons but lives on a soliton configuration. Perhaps
the simplest example of this state is the localized excitation of the kink
solution in $\lambda \phi^4$ theory. One can show that at a linear
level there is such a bound state associated with the width of the kink
solution and whose frequency is below the lowest frequency states in the vacuum. 
This means that, as in the case of oscillons, their decay is due to  
non-linear interactions.

In this paper we have studied the full non-linear evolution of such
localized bound state of the kink soliton in Minkowski space. We have been able to do
this over long periods of time by using a $(1+1)$ dimensional
lattice field theory with absorbing boundary conditions. We have shown
that one could store quite a large portion of extra energy on this bound
state for times much longer than the natural time scale associated with
their width. The similarities with the breather solutions in the theory
makes us think of these bound states on the kink as a breather
trapped in the core of the soliton.

We have also explored the evolution of these kink solutions
in several cosmological backgrounds. In particular, we have run a
large number of realizations to simulate a cosmological phase
transition that leads to the formation of these defects. We found that
the solitons will typically get formed with an approximately $20\%$ 
of extra energy due to the presence of a substantial amplitude of the
shape mode.

We have also simulated the interaction of the kink solution with
a thermal bath and computed the average value of the expected amplitude
of the bound state as a function of the temperature. As one
increases the thermal energy of the background, the amplitude
of the bound state grows following the relation $\langle \hat A \rangle \propto T^{1/2}$.
This amplitude saturates when the extra energy of the soliton is also
about $20\%$. This suggests that a purely thermal formation
of solitons would also create them with some extra energy
at this $20\%$ bound.

Our results suggest that, in a realistic setting, solitons
will always be formed in an excited state with some extra energy
stored in their bound state modes. This could have important consequences for the subsequent
evolution of these objects. In particular, if this type of
behaviour persists in higher dimensions, adding some extra
energy to the solitonic configurations can easily affect their
equation of state and hence their dynamics.

It is then clear that all these effects could modify the predictions of field theory
simulations of these solitons since they could remain excited
for the duration of the numerical span of the run. However, even
though the lifetime of the bound states is very long compared
to any time scale of the simulation, it is many orders of
magnitude smaller than any relevant cosmological time. This
means that one should also investigate whether there
is any mechanism that maintains the level of excitation
of the solitons throughout their cosmological evolution.
Therefore, it remains to be seen if this initial energy that
we report in this paper has a cosmological relevance or
is just a short transient effect. We are currently investigating
these ideas and we hope to report on them in the near future.

\section{Acknowledgements}

We are grateful to Ken Olum, Jose Queiruga, Tanmay Vachaspati, Manuel Valle and Alex Vilenkin for very useful 
suggestions and discussions.  This work is supported in part by the Spanish Ministry MCIU/AEI/FEDER 
grant (PGC2018-094626-B-C21), the Basque Government grant (IT-979-16) and the Basque 
Foundation for Science (IKERBASQUE). The numerical work carried out in this paper has been 
possible thanks to the computing infrastructure of the ARINA cluster at the University of the 
Basque Country, (UPV/EHU), and the use of the Harria computer. We are grateful to 
Mariam Bouhmadi-Lopez for providing us access to Harria where some of this calculation 
were run.

\bibliography{kinkexcitations}

\appendix
\section{Numerical details}
\label{numericaldetails}
\subsection{Discrete Lattice evolution equations}
\label{discretization}

We wish to solve numerically the following equations of motion (\ref{eomback}):
\beq
\ddot \phi + H \dot \phi - \frac{1}{a^2} \phi'' + \lambda \phi \left(\phi^2 - \eta^2\right)= 0~,
\eeq
where dots and primes denote derivatives with respect to cosmic time and comoving space
respectively, and $H = \dot a /a$ is the Hubble rate.
For the numerical discretization of the equations of motion, first we revert to dimensionless quantities
\beq
\tilde \phi = \frac{\phi}{\eta} ~~~~~~~~~~~~~~ \tilde x = \sqrt{\lambda} \eta x~~~~~~~~~~~~~~\tilde t = \sqrt{\lambda} \eta t~~,
\eeq
and dots and primes will now mean derivatives with respect to dimensionless quantities.
The equation of motion now reads
\begin{equation}
\ddot{\tilde{\phi}}-\frac{1}{a^{2}}\tilde{\phi}^{\prime\prime}+\tilde\phi \left(\tilde\phi^2 - 1\right)+\tilde{H}\dot{\tilde{\phi}}=0\,,
\label{eomdisc}
\eeq
where now $\tilde{H}=\frac{H}{\sqrt{\lambda}\eta}$ is the dimensionless Hubble rate. If the evolution 
takes place in Minkowski spacetime, it suffices to set $a=1$ and $\tilde{H}=0$.
(From now on we will drop the tildes for simplicity)

We need to discretize equation~(\ref{eomdisc}) into the lattice. In order to discretize the time derivatives, 
we use the so-called  staggered leapfrog method. In order to do that,  we need to define the conjugate momentum as
\begin{equation}
 {\pi}\!\left( {x}, {t}+\frac{\Delta {t}}{2}\right)
\equiv\frac{ {\phi}\left( {x}, {t}+\Delta {t}\right)- {\phi}\left( {x}, {t}\right)}{\Delta {t}}\,. \label{eq:staggered leapfrog method}
\end{equation} 
Note that the field lives in integer time-steps, and the conjugate momentum in half time-steps.
Now, the equation of motion for $\pi$ reads 
\beq 
\dot{ {\pi}}=\frac{1}{a^{2}} {\phi}^{\prime\prime}-\phi \left(\phi^2 - 1\right)- {H} {\pi}\,.
\label{eq:eom field plus friction}
\end{equation}
First we discretize the second  order spatial derivative of the field  using nearest neighbours:
\begin{equation}
{\phi}''\left( {x}, {t}\right)=\frac{ {\phi}\left( {x}+\Delta {x}, {t}\right)-2 {\phi}\left( {x}, {t}\right)+ {\phi}\left( {x}-\Delta {x}, {t}\right)}{\left(\Delta {x}\right)^{2}}\,\,.
\label{eq:second derivative}
\end{equation}
Since $\dot{{\pi}}$ will be evaluated at an integer time step as it is calculated 
as the difference of $ {\pi}$ at two different half-integer time steps:
\begin{equation}
\dot{ {\pi}}\left( {x}, {t}+\Delta {t}\right)=\frac{ {\pi}\left( {x}, {t}+\frac{3}{2}\Delta {t}\right)- {\pi}\left( {x}, {t}+\frac{\Delta {t}}{2}\right)}{\Delta {t}}.
\label{eq:staggered leapfrog method pi dot}
\end{equation}
From (\ref{eq:eom field  plus friction}) we see that $ {\pi}$ and $\dot{ {\pi}}$ should 
be evaluated at the same time step. However, the former lives at half-integer steps, while 
the latter lives at integer steps. To solve this issue, we simply replace $\pi$  in the friction term with 
its average in neighbouring half time-steps:
\begin{equation}
\pi(x,t+\Delta t)=\frac{ {\pi}\left( {x}, {t}+\frac{1}{2}\Delta {t}\right)+ {\pi}\left( {x}, {t}+\frac{3}{2}\Delta {t}\right)}{2}\,\,.
\label{eq:staggered leapfrog method viscosity term}
\end{equation}
Now all the terms in the equation for $\pi$ (\ref{eq:eom field plus friction}) are evaluated at integer 
time-steps. The staggered leapfrog method now consists of solving for the conjugate momentum 
using (\ref{eq:eom field plus friction}), which explicitly reads
\begin{eqnarray}
  {\pi}( {x}, {t}+\frac{3}{2}\Delta {t})&=&\left(1+H\frac{ \Delta {t}}{2}\right)^{-1}\left\{\left(1-H\frac{ \Delta {t}}{2}\right) {\pi}( {x}, {t}+\frac{1}{2}\Delta {t} )+\right.\nonumber\\
&&\left.+ \Delta {t}\left[ \frac{1}{a(t)^2} \phi''({x}, {t})-\phi(\phi^2-1)\right]\right\}\,,
\label{eq:staggered leapfrog method pi}
\end{eqnarray}
with $\phi''$ given by eq.~(\ref{eq:second derivative}), and then solving for the field, which can be obtained from eq.~(\ref{eq:staggered leapfrog method}):
\begin{equation}
 {\phi}( {x}, {t}+\Delta {t})= {\phi}\left( {x}, {t}\right)+\Delta {t}\,\, \pi\!\left( {x}, {t}+\frac{\Delta {t}}{2}\right).
\label{eq:staggered leapfrog method phi}
\end{equation}

The simulations are in 1-dimensional space, so in principle they are not very demanding, 
computer-wise. However, some of them have to be performed for very long periods of time. Moreover, 
since our simulation box is expanding in comoving coordinates, and the physical size of our 
objects is fixed, we encounter the problem that by the end of the simulation we may not have 
enough points to resolve out objects (the kink or the shape mode, e.g.) satisfactorily.

We thus wrote a parallel version of the code (also anticipating our higher dimensional simulations). 
The simulated (1-dimensional) box is distributed among different processors. Since our finite-difference 
scheme only relies on nearest neighbours, we only need a one dimensional halo around the box in 
each processor. We used {\it message passing interface} (MPI)  for communication between 
different   processors. A typical processor will share its information with the processors to either side 
of it. Except the first and last processors, which only share their information with one processor (to 
their right and left, respectively). On the other side of these processors, we apply  the boundary 
conditions, absorbing boundary conditions, which are described below.

 A typical simulation consists of a box of length $L=50$, with $\Delta x=0.01$ and $\Delta t=0.0008$. Thus, 
 the number of points in the box is $N=5000$, which are distributed among 40 processors.

\subsection{Absorbing Boundary Conditions}
\label{AbsorbingBC}

The idea behind these boundary conditions is quite simply to force an outgoing wave traveling towards
the boundary to be annihilated at that point \cite{ABC} . These conditions read
\begin{equation}
\left(\partial_{x}\phi+\partial_{t}\phi\right)\bigg\rvert_{x=\frac{L}{2},t}=0,
\label{eq:right boundary}
\end{equation}

\begin{equation}
\left(\partial_{x}\phi-\partial_{t}\phi\right)\bigg\rvert_{x=-\frac{L}{2},t}=0.
\label{eq:left boundary}
\end{equation}

At the boundaries,  we will assume that the field can be written as a small perturbation around one of the 
vacua, namely, $\phi=\eta+\xi$. Let us consider a linear perturbation traveling, for instance, towards the right 
boundary. The solution is given by
\begin{equation}
\xi\left(x,t\right)\propto e^{i\left(kx-\omega t\right)}.
\label{eq:plane wave}
\end{equation}

The equation of motion for the perturbation, at $\mathcal{O}\left(\xi\right)$, is
\begin{equation}
\ddot{\xi}-\xi^{''}+m^{2}\xi=0\,.
\label{eq:perturbations about vacuum}
\end{equation}

The travelling wave (\ref{eq:plane wave}) is a solution to the equation of motion 
if $\omega^{2}=k^{2}+m^{2}$, but it only solves the absorbing boundary conditions if $\omega=k$. This means 
that these conditions will work perfectly for those modes with $k\gg m$, in other words, for the relativistic modes.

It is possible that the outgoing radiation in a particular problem is monochromatic, with known angular 
frequency $\omega$. For instance, if our initial condition consists of a kink with its shape mode excited, we 
know that it will radiate waves with frequency $\omega=2\,\omega_{s}=\sqrt{3}\,m$. In such a situation, the 
absorbing boundary conditions can be refined as follows:
\begin{equation}
\left(\frac{\omega}{\sqrt{\omega^{2}-m^{2}}}\,\partial_{x}\phi+\partial_{t}\phi\right)\bigg\rvert_{x=\frac{L}{2},t}=0\,,
\label{eq:mur right}
\end{equation}
\begin{equation}
\left(\frac{\omega}{\sqrt{\omega^{2}-m^{2}}}\,\partial_{x}\phi-\partial_{t}\phi\right)\bigg\rvert_{x=-\frac{L}{2},t}=0\,.
\label{eq:mur left}
\end{equation}
One can easily check that a traveling wave satisfying the equation of motion also satisfies these two equations.
This means that for our almost monochromatic radiation this boundary conditions will be more
effective. We have implemented these conditions and indeed the absorption gets to be 
better with this modified absorbing boundary conditions.

Here we show a couple of different situations to illustrate the way this
absorbing boundary conditions work. We run our code twice in flat space with an initial
condition given by an excited kink, but the excitations in both cases are
quite different. In the first case, we use the bound state solution as the excitation
with amplitude $A(0)= 0.332$ . In the second case, we use a symmetric perturbation
with the same amount of energy. 

\begin{figure}[h!]
\includegraphics[width=15.6cm]{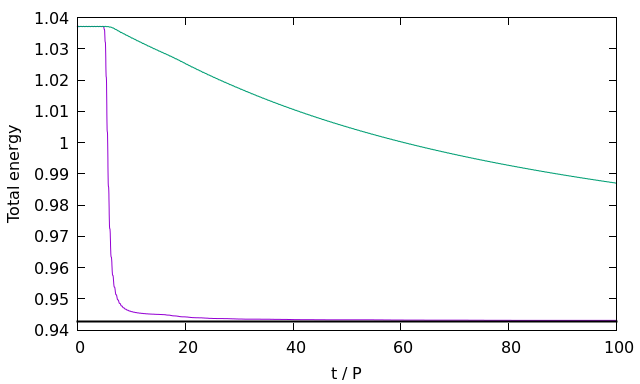}
\label{fig:bc-at-work}
\caption{Total energy inside of the box for two different initial conditions with 
absorbing boundary conditions. The green line shows the slow decay of the 
energy for a kink with a bound state excitation. The purple line represents the
configuration of a kink with a symmetric perturbation. The total energy in this case
is very quickly emitted from the kink and absorbed almost immediately as it reaches
the boundaries.}
\end{figure}

We plot in Fig.~\ref{fig:bc-at-work} the energy inside of our simulation box over the simulation time. 
We use the same boundary conditions in both cases. The results indicate several
important points. First, the energy remains perfectly constant in both cases 
before any perturbation reaches the boundary. This means our code conserves
energy in Minkowski space as it should. The energy curve in both cases starts
to decline as soon as the first waves arrive to the boundary. However, the
total energy from the symmetric perturbation decreases quite rapidly and
in a short period of time the final energy is the one of the kink itself. That means
that the generic absorbing conditions work quite efficiently and there is not
much energy bouncing around the box. In the other case, the energy curve
decreases slowly due to the presence of the bound state that decays slowly. This slow leakage of energy is also efficiently absorbed in the 
boundary and the final result is also a small amount of feedback into
the bound state due to energy bouncing off the edges of the box.

\subsection{Metropolis Algorithm for a thermal state}
\label{Metropolis}

In the main part of the text we make use of a thermal state for a free massive
scalar field as the initial conditions for some of our simulations. We do this by
initializing the power spectrum of the field and its momentum so that their 
occupation numbers satisfy the Bose-Einstein distribution. 

Here, we use an alternative way to arrive at this state that is based on the
classic Metropolis algorithm \cite{Metropolis}. This procedure is more
flexible since it does not assume that we are in a free field vacuum.
In fact, we will also use it to consider thermal fluctuations
around the background kink solution, which may seem to be a 
more accurate method than the one used in the main text.

\begin{figure}[h!]
\begingroup
  \makeatletter
  \providecommand\color[2][]{%
    \GenericError{(gnuplot) \space\space\space\@spaces}{%
      Package color not loaded in conjunction with
      terminal option `colourtext'%
    }{See the gnuplot documentation for explanation.%
    }{Either use 'blacktext' in gnuplot or load the package
      color.sty in LaTeX.}%
    \renewcommand\color[2][]{}%
  }%
  \providecommand\includegraphics[2][]{%
    \GenericError{(gnuplot) \space\space\space\@spaces}{%
      Package graphicx or graphics not loaded%
    }{See the gnuplot documentation for explanation.%
    }{The gnuplot epslatex terminal needs graphicx.sty or graphics.sty.}%
    \renewcommand\includegraphics[2][]{}%
  }%
  \providecommand\rotatebox[2]{#2}%
  \@ifundefined{ifGPcolor}{%
    \newif\ifGPcolor
    \GPcolorfalse
  }{}%
  \@ifundefined{ifGPblacktext}{%
    \newif\ifGPblacktext
    \GPblacktexttrue
  }{}%
  \let\gplgaddtomacro\g@addto@macro
  \gdef\gplbacktext{}%
  \gdef\gplfronttext{}%
  \makeatother
  \ifGPblacktext
    \def\colorrgb#1{}%
    \def\colorgray#1{}%
  \else
    \ifGPcolor
      \def\colorrgb#1{\color[rgb]{#1}}%
      \def\colorgray#1{\color[gray]{#1}}%
      \expandafter\def\csname LTw\endcsname{\color{white}}%
      \expandafter\def\csname LTb\endcsname{\color{black}}%
      \expandafter\def\csname LTa\endcsname{\color{black}}%
      \expandafter\def\csname LT0\endcsname{\color[rgb]{1,0,0}}%
      \expandafter\def\csname LT1\endcsname{\color[rgb]{0,1,0}}%
      \expandafter\def\csname LT2\endcsname{\color[rgb]{0,0,1}}%
      \expandafter\def\csname LT3\endcsname{\color[rgb]{1,0,1}}%
      \expandafter\def\csname LT4\endcsname{\color[rgb]{0,1,1}}%
      \expandafter\def\csname LT5\endcsname{\color[rgb]{1,1,0}}%
      \expandafter\def\csname LT6\endcsname{\color[rgb]{0,0,0}}%
      \expandafter\def\csname LT7\endcsname{\color[rgb]{1,0.3,0}}%
      \expandafter\def\csname LT8\endcsname{\color[rgb]{0.5,0.5,0.5}}%
    \else
      \def\colorrgb#1{\color{black}}%
      \def\colorgray#1{\color[gray]{#1}}%
      \expandafter\def\csname LTw\endcsname{\color{white}}%
      \expandafter\def\csname LTb\endcsname{\color{black}}%
      \expandafter\def\csname LTa\endcsname{\color{black}}%
      \expandafter\def\csname LT0\endcsname{\color{black}}%
      \expandafter\def\csname LT1\endcsname{\color{black}}%
      \expandafter\def\csname LT2\endcsname{\color{black}}%
      \expandafter\def\csname LT3\endcsname{\color{black}}%
      \expandafter\def\csname LT4\endcsname{\color{black}}%
      \expandafter\def\csname LT5\endcsname{\color{black}}%
      \expandafter\def\csname LT6\endcsname{\color{black}}%
      \expandafter\def\csname LT7\endcsname{\color{black}}%
      \expandafter\def\csname LT8\endcsname{\color{black}}%
    \fi
  \fi
    \setlength{\unitlength}{0.0500bp}%
    \ifx\gptboxheight\undefined%
      \newlength{\gptboxheight}%
      \newlength{\gptboxwidth}%
      \newsavebox{\gptboxtext}%
    \fi%
    \setlength{\fboxrule}{0.5pt}%
    \setlength{\fboxsep}{1pt}%
\begin{picture}(8844.00,5668.00)%
    \gplgaddtomacro\gplbacktext{%
      \csname LTb\endcsname
      \put(814,704){\makebox(0,0)[r]{\strut{}$0$}}%
      \put(814,1653){\makebox(0,0)[r]{\strut{}$20$}}%
      \put(814,2601){\makebox(0,0)[r]{\strut{}$40$}}%
      \put(814,3550){\makebox(0,0)[r]{\strut{}$60$}}%
      \put(814,4498){\makebox(0,0)[r]{\strut{}$80$}}%
      \put(814,5447){\makebox(0,0)[r]{\strut{}$100$}}%
      \put(946,484){\makebox(0,0){\strut{}$0$}}%
      \put(2446,484){\makebox(0,0){\strut{}$20$}}%
      \put(3946,484){\makebox(0,0){\strut{}$40$}}%
      \put(5447,484){\makebox(0,0){\strut{}$60$}}%
      \put(6947,484){\makebox(0,0){\strut{}$80$}}%
      \put(8447,484){\makebox(0,0){\strut{}$100$}}%
    }%
    \gplgaddtomacro\gplfronttext{%
      \csname LTb\endcsname
      \put(198,3075){\rotatebox{-270}{\makebox(0,0){\Large Energy}}}%
      \put(4696,154){\makebox(0,0){\Large Number of sweeps ($\times 10^{4}$)}}%
      \csname LTb\endcsname
      \put(7460,5274){\makebox(0,0)[r]{\strut{}Kinetic energy}}%
      \csname LTb\endcsname
      \put(7460,5054){\makebox(0,0)[r]{\strut{}Gradient $+$ Potential energy}}%
      \csname LTb\endcsname
      \put(7460,4834){\makebox(0,0)[r]{\strut{} Total energy}}%
      \csname LTb\endcsname
      \put(7460,4614){\makebox(0,0)[r]{\strut{}$\tilde{M}_{k}+N\theta$}}%
      \csname LTb\endcsname
      \put(7460,4394){\makebox(0,0)[r]{\strut{}$ N\theta/2$}}%
      \csname LTb\endcsname
      \put(7460,4174){\makebox(0,0)[r]{\strut{}$\tilde{M}_{k}+N\theta/2$}}%
    }%
    \gplbacktext
    \put(0,0){\includegraphics{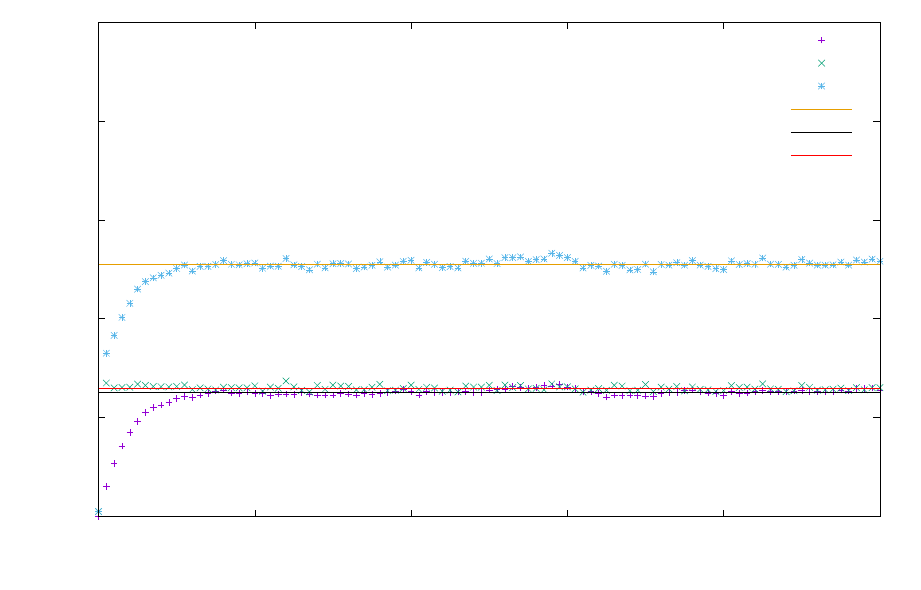}}%
    \gplfronttext
  \end{picture}%
\endgroup
\caption{Energy throughout the Metropolis algorithm (example with $\theta=0.01$), taking the kink solution as initial state. 
The kinetic energy tends to $N\theta/2$, while the gradient+potential energy 
saturates to $\tilde{M}_{k}+N\theta/2$.}
\label{fig:energy metropolis}
\end{figure}

\begin{figure}[h!]
\begingroup
  \makeatletter
  \providecommand\color[2][]{%
    \GenericError{(gnuplot) \space\space\space\@spaces}{%
      Package color not loaded in conjunction with
      terminal option `colourtext'%
    }{See the gnuplot documentation for explanation.%
    }{Either use 'blacktext' in gnuplot or load the package
      color.sty in LaTeX.}%
    \renewcommand\color[2][]{}%
  }%
  \providecommand\includegraphics[2][]{%
    \GenericError{(gnuplot) \space\space\space\@spaces}{%
      Package graphicx or graphics not loaded%
    }{See the gnuplot documentation for explanation.%
    }{The gnuplot epslatex terminal needs graphicx.sty or graphics.sty.}%
    \renewcommand\includegraphics[2][]{}%
  }%
  \providecommand\rotatebox[2]{#2}%
  \@ifundefined{ifGPcolor}{%
    \newif\ifGPcolor
    \GPcolorfalse
  }{}%
  \@ifundefined{ifGPblacktext}{%
    \newif\ifGPblacktext
    \GPblacktexttrue
  }{}%
  \let\gplgaddtomacro\g@addto@macro
  \gdef\gplbacktext{}%
  \gdef\gplfronttext{}%
  \makeatother
  \ifGPblacktext
    \def\colorrgb#1{}%
    \def\colorgray#1{}%
  \else
    \ifGPcolor
      \def\colorrgb#1{\color[rgb]{#1}}%
      \def\colorgray#1{\color[gray]{#1}}%
      \expandafter\def\csname LTw\endcsname{\color{white}}%
      \expandafter\def\csname LTb\endcsname{\color{black}}%
      \expandafter\def\csname LTa\endcsname{\color{black}}%
      \expandafter\def\csname LT0\endcsname{\color[rgb]{1,0,0}}%
      \expandafter\def\csname LT1\endcsname{\color[rgb]{0,1,0}}%
      \expandafter\def\csname LT2\endcsname{\color[rgb]{0,0,1}}%
      \expandafter\def\csname LT3\endcsname{\color[rgb]{1,0,1}}%
      \expandafter\def\csname LT4\endcsname{\color[rgb]{0,1,1}}%
      \expandafter\def\csname LT5\endcsname{\color[rgb]{1,1,0}}%
      \expandafter\def\csname LT6\endcsname{\color[rgb]{0,0,0}}%
      \expandafter\def\csname LT7\endcsname{\color[rgb]{1,0.3,0}}%
      \expandafter\def\csname LT8\endcsname{\color[rgb]{0.5,0.5,0.5}}%
    \else
      \def\colorrgb#1{\color{black}}%
      \def\colorgray#1{\color[gray]{#1}}%
      \expandafter\def\csname LTw\endcsname{\color{white}}%
      \expandafter\def\csname LTb\endcsname{\color{black}}%
      \expandafter\def\csname LTa\endcsname{\color{black}}%
      \expandafter\def\csname LT0\endcsname{\color{black}}%
      \expandafter\def\csname LT1\endcsname{\color{black}}%
      \expandafter\def\csname LT2\endcsname{\color{black}}%
      \expandafter\def\csname LT3\endcsname{\color{black}}%
      \expandafter\def\csname LT4\endcsname{\color{black}}%
      \expandafter\def\csname LT5\endcsname{\color{black}}%
      \expandafter\def\csname LT6\endcsname{\color{black}}%
      \expandafter\def\csname LT7\endcsname{\color{black}}%
      \expandafter\def\csname LT8\endcsname{\color{black}}%
    \fi
  \fi
    \setlength{\unitlength}{0.0500bp}%
    \ifx\gptboxheight\undefined%
      \newlength{\gptboxheight}%
      \newlength{\gptboxwidth}%
      \newsavebox{\gptboxtext}%
    \fi%
    \setlength{\fboxrule}{0.5pt}%
    \setlength{\fboxsep}{1pt}%
\begin{picture}(8844.00,5668.00)%
    \gplgaddtomacro\gplbacktext{%
      \csname LTb\endcsname
      \put(1342,704){\makebox(0,0)[r]{\strut{}$ 10^{-\-8}$}}%
      \put(1342,1382){\makebox(0,0)[r]{\strut{}$ 10^{-\-7}$}}%
      \put(1342,2059){\makebox(0,0)[r]{\strut{}$ 10^{-\-6}$}}%
      \put(1342,2737){\makebox(0,0)[r]{\strut{}$ 10^{-\-5}$}}%
      \put(1342,3414){\makebox(0,0)[r]{\strut{}$ 10^{-\-4}$}}%
      \put(1342,4092){\makebox(0,0)[r]{\strut{}$ 10^{-\-3}$}}%
      \put(1342,4769){\makebox(0,0)[r]{\strut{}$ 10^{-\-2}$}}%
      \put(1342,5447){\makebox(0,0)[r]{\strut{}$ 10^{-\-1}$}}%
      \put(3323,484){\makebox(0,0){\strut{}$1$}}%
      \put(5375,484){\makebox(0,0){\strut{}$10$}}%
      \put(7427,484){\makebox(0,0){\strut{}$100$}}%
    }%
    \gplgaddtomacro\gplfronttext{%
      \csname LTb\endcsname
      \put(198,3075){\rotatebox{-270}{\makebox(0,0){\Large $\langle|\tilde{\xi}_{n}|^{2}\rangle$}}}%
      \put(4960,154){\makebox(0,0){\Large $\tilde{k}_{n}$}}%
    }%
    \gplbacktext
    \put(0,0){\includegraphics{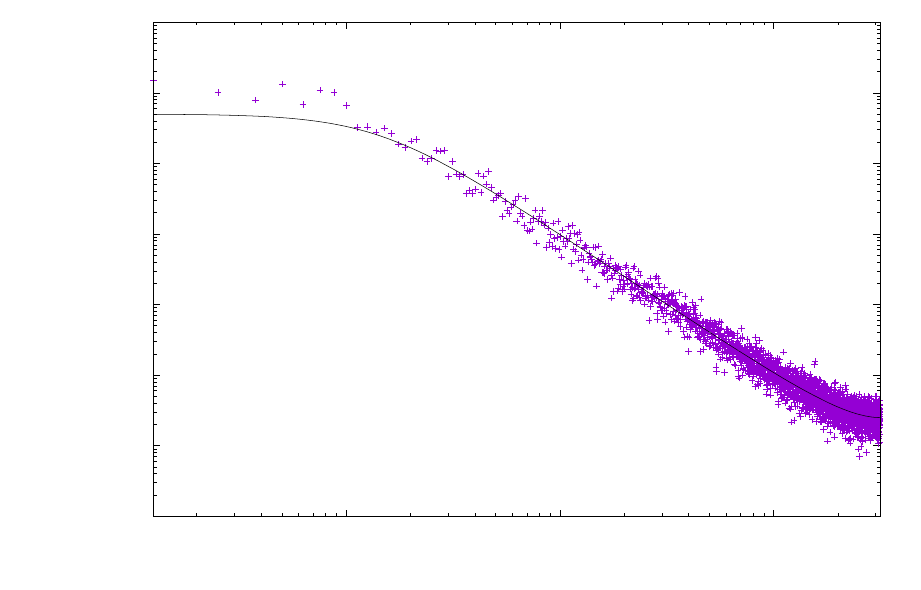}}%
    \gplfronttext
  \end{picture}%
\endgroup
\caption{Fourier spectrum of the thermal fluctuations $\tilde{\xi}$ at temperature
 $\theta=0.01$ (average of 10 Metropolis realizations). The black curve corresponds
  to the classical limit of the theoretical spectrum at this temperature.}
\label{fig:field-spectrum-metropolis}
\end{figure}

Let us briefly review the main steps to achieve this thermal state.
The initial field profile for the algorithm is chosen to be 
a constant value at one of the vacua\footnote{For the thermal state around the
kink solution we start with the kink solution instead.} and we 
set the initial field velocity to zero. The algorithm then
consists of the following steps:
\begin{enumerate}
\item We generate a new configuration which differs from the present one by 
changing the value of the field and its velocity at one lattice point\footnote{Since field 
and field velocity are independent variables, they must be varied  
independently.}.
\item We then calculate the difference in energy between the new state 
and the old one: $\Delta\tilde{E}=\tilde{E}_{new}-\tilde{E}_{old}$. Now there are two possibilities: 
\begin{enumerate}
\item The energy has decreased or remained the same: $\Delta\tilde{E}\leq0$. In this case, we accept the change.
\item The energy has increased: $\Delta\tilde{E}>0$. In this case, we accept the change with probability 
\begin{equation}
p=e^{-\frac{\Delta\tilde{E}}{\theta}}~,
\label{eq:acceptance probability}
\end{equation} 
with $\theta$ the temperature at which we are simulating.
This can be done as follows. We choose a random number $r$ between zero and one, $0\leq r<1$. If that 
number is less than the acceptance probability, $r<p$, then we accept the change. Otherwise, we leave the 
value of the field and the field velocity unchanged.
\end{enumerate}
\end{enumerate} 

After that we just have to follow the same steps over and over again until the field reaches thermal equilibrium. We 
will consider that the field is in a state of thermal equilibrium when the following two conditions are satisfied: 
\begin{enumerate}
\item The total energy of the perturbations has saturated to $N\theta$, and it is equally stored in kinetic 
energy and gradient$+$potential energy, in agreement with equipartition (see figure \ref{fig:energy metropolis}).
\item The Fourier spectrum of both the field and the field velocity perturbations are well approximated by 
the classical limit of the theoretical thermal spectrum described in Eq. (\ref{eq:occupation number tic}).
\end{enumerate}

\begin{figure}[h!]
\begingroup
  \makeatletter
  \providecommand\color[2][]{%
    \GenericError{(gnuplot) \space\space\space\@spaces}{%
      Package color not loaded in conjunction with
      terminal option `colourtext'%
    }{See the gnuplot documentation for explanation.%
    }{Either use 'blacktext' in gnuplot or load the package
      color.sty in LaTeX.}%
    \renewcommand\color[2][]{}%
  }%
  \providecommand\includegraphics[2][]{%
    \GenericError{(gnuplot) \space\space\space\@spaces}{%
      Package graphicx or graphics not loaded%
    }{See the gnuplot documentation for explanation.%
    }{The gnuplot epslatex terminal needs graphicx.sty or graphics.sty.}%
    \renewcommand\includegraphics[2][]{}%
  }%
  \providecommand\rotatebox[2]{#2}%
  \@ifundefined{ifGPcolor}{%
    \newif\ifGPcolor
    \GPcolorfalse
  }{}%
  \@ifundefined{ifGPblacktext}{%
    \newif\ifGPblacktext
    \GPblacktexttrue
  }{}%
  \let\gplgaddtomacro\g@addto@macro
  \gdef\gplbacktext{}%
  \gdef\gplfronttext{}%
  \makeatother
  \ifGPblacktext
    \def\colorrgb#1{}%
    \def\colorgray#1{}%
  \else
    \ifGPcolor
      \def\colorrgb#1{\color[rgb]{#1}}%
      \def\colorgray#1{\color[gray]{#1}}%
      \expandafter\def\csname LTw\endcsname{\color{white}}%
      \expandafter\def\csname LTb\endcsname{\color{black}}%
      \expandafter\def\csname LTa\endcsname{\color{black}}%
      \expandafter\def\csname LT0\endcsname{\color[rgb]{1,0,0}}%
      \expandafter\def\csname LT1\endcsname{\color[rgb]{0,1,0}}%
      \expandafter\def\csname LT2\endcsname{\color[rgb]{0,0,1}}%
      \expandafter\def\csname LT3\endcsname{\color[rgb]{1,0,1}}%
      \expandafter\def\csname LT4\endcsname{\color[rgb]{0,1,1}}%
      \expandafter\def\csname LT5\endcsname{\color[rgb]{1,1,0}}%
      \expandafter\def\csname LT6\endcsname{\color[rgb]{0,0,0}}%
      \expandafter\def\csname LT7\endcsname{\color[rgb]{1,0.3,0}}%
      \expandafter\def\csname LT8\endcsname{\color[rgb]{0.5,0.5,0.5}}%
    \else
      \def\colorrgb#1{\color{black}}%
      \def\colorgray#1{\color[gray]{#1}}%
      \expandafter\def\csname LTw\endcsname{\color{white}}%
      \expandafter\def\csname LTb\endcsname{\color{black}}%
      \expandafter\def\csname LTa\endcsname{\color{black}}%
      \expandafter\def\csname LT0\endcsname{\color{black}}%
      \expandafter\def\csname LT1\endcsname{\color{black}}%
      \expandafter\def\csname LT2\endcsname{\color{black}}%
      \expandafter\def\csname LT3\endcsname{\color{black}}%
      \expandafter\def\csname LT4\endcsname{\color{black}}%
      \expandafter\def\csname LT5\endcsname{\color{black}}%
      \expandafter\def\csname LT6\endcsname{\color{black}}%
      \expandafter\def\csname LT7\endcsname{\color{black}}%
      \expandafter\def\csname LT8\endcsname{\color{black}}%
    \fi
  \fi
    \setlength{\unitlength}{0.0500bp}%
    \ifx\gptboxheight\undefined%
      \newlength{\gptboxheight}%
      \newlength{\gptboxwidth}%
      \newsavebox{\gptboxtext}%
    \fi%
    \setlength{\fboxrule}{0.5pt}%
    \setlength{\fboxsep}{1pt}%
\begin{picture}(8844.00,5668.00)%
    \gplgaddtomacro\gplbacktext{%
      \csname LTb\endcsname
      \put(1078,704){\makebox(0,0)[r]{\strut{}$0.001$}}%
      \put(1078,3076){\makebox(0,0)[r]{\strut{}$0.01$}}%
      \put(1078,5447){\makebox(0,0)[r]{\strut{}$0.1$}}%
      \put(3129,484){\makebox(0,0){\strut{}$1$}}%
      \put(5258,484){\makebox(0,0){\strut{}$10$}}%
      \put(7388,484){\makebox(0,0){\strut{}$100$}}%
    }%
    \gplgaddtomacro\gplfronttext{%
      \csname LTb\endcsname
      \put(198,3075){\rotatebox{-270}{\makebox(0,0){\Large $\langle|\tilde{\pi}_{n}|^{2}\rangle$}}}%
      \put(4828,154){\makebox(0,0){\Large $\tilde{k}_{n}$}}%
    }%
    \gplbacktext
    \put(0,0){\includegraphics{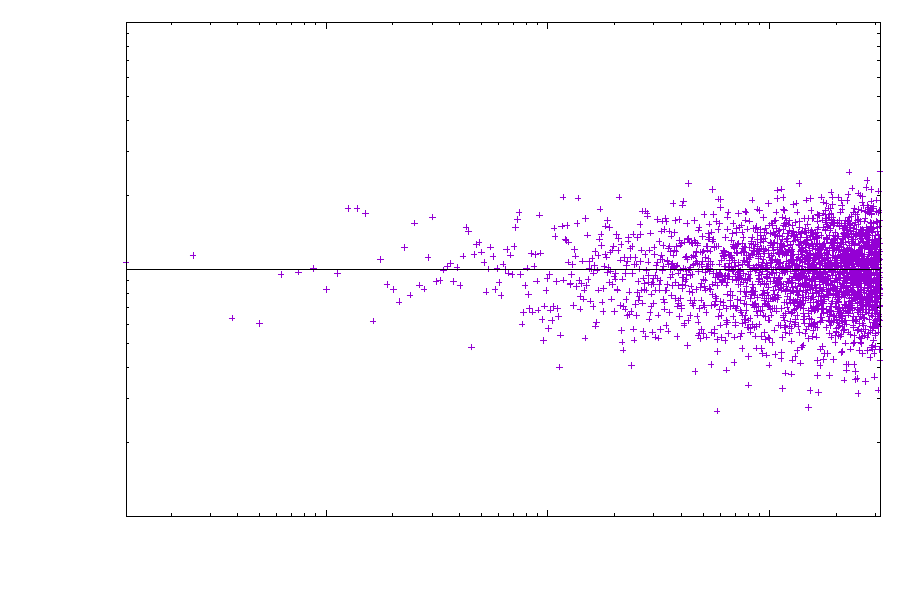}}%
    \gplfronttext
  \end{picture}%
\endgroup
\caption{Fourier spectrum of the velocity thermal fluctuations $\tilde{\pi}$ at temperature 
$\theta=0.01$ (average of 10 Metropolis realizations). The black curve corresponds to the 
classical limit of the theoretical spectrum at this temperature.}
\label{fig:velocity-spectrum-metropolis}
\end{figure}

We show in Figs. (\ref{fig:field-spectrum-metropolis}) and (\ref{fig:velocity-spectrum-metropolis}) the
comparison of the different spectra obtained using the Metropolis algorithm and 
the classical limit of the theoretical spectrum at the same temperature. The coefficients
$\tilde \xi_n$  
are the Fourier coefficients of the thermal perturbation field, and they are related to the $\tilde\alpha_{n}$ parameters used in the main text by
\beq
\tilde \xi_{n} = \frac{1}{\sqrt{2\tilde \omega_{n}}}\left(\tilde \alpha_{n}+\tilde \alpha_{-n}^{*}\right)~.
\label{xi-alpha}
\eeq

The $\tilde \pi_n$ are the analogous coefficients
for the field velocity:
\beq
\tilde \pi_{n} = i\sqrt{\frac{\tilde \omega_{n}}{2}}\left(-\tilde \alpha_{n}+\tilde \alpha_{-n}^{*}\right)~.
\label{pi-alpha}
\eeq

We have also used the Metropolis algorithm to generate the spectrum of excitations of a
kink in thermal equilibrium with a background. The simulations performed with
these initial conditions yield statistically indistinguishable results from the 
classical limit of the spectrum Eq. (\ref{eq:tic phi phys}).  However, this method is much more 
costly in terms of time. This is why at the end we use the other method to generate a large
set of realizations.

\section{Analytical estimate of the decay rate of the bound state}
\label{analytic-decay}

As seen in the main text, at the linear level, the bound state (\ref{eq:bound-state-mode})
should oscillate with frequency $w_s = \sqrt{\frac{3}{2}}$ and should not decay, since its 
frequency is smaller than the frequencies of the propagating modes outside the kink. 
However, since the bound state couples nonlinearly to the propagating 
modes, the energy of the bound state is radiated away.
As a consequence, the oscillating amplitude of the shape mode slowly decreases 
with time. The particular time dependence of the amplitude can be estimated analytically. Let 
us describe this estimation following the calculation given in \cite{Manton:1996ex}, where we have adapted
it to our notation.

We consider a parametrization of the field given by
\beq
\phi(x,t) = \phi_k(x) + A(t) \bar f_s(x) + f_r(x,t)~,
\label{decomposition}
\eeq
where the $f_r(x,t)$ term is related to the propagating modes (\ref{eq:propagating}), 
and as said before, they are orthogonal to $\bar f_s$
\beq
\int {\rm d}x \bar f_s(x) f_r(x,t) = 0 \,.
\eeq

Plugging the decomposition in (\ref{decomposition})  into the equation of motion, one can see 
that at the lowest order in $A$, the frequency of the oscillation is $w_s$, as we mentioned 
before. At the next order, at $O(A^2)$, the equation becomes,
\begin{equation}
\left(\ddot{A}+\frac{3}{2}A\right)\bar f_{s}+\ddot{f}_{r}-f_{r}^{''}+\left(3\phi_{k}^{2}-1\right)f_{r}=-3\phi_{k}\bar f_{s}^{2}A^{2}.
\label{eq:asm second order}
\end{equation}

We can now multiply this equation by the shape mode function $\bar f_s(x)$ and integrate over 
all space, employing the orthogonality of eigenstates. This operation yields \footnote{Note that 
according to the equation of linear perturbations, the term $-f_{r}^{''}+(3\phi_{k}^{2}-1)f_{r}$ is 
proportional to $f_{r}$, and therefore it is orthogonal to the shape mode.}
\begin{equation}
\ddot{A}+\frac{3}{2}A=-3\alpha A^{2},
\label{eq:second order one}
\end{equation}
where
\begin{equation}
\alpha=\int_{-\infty}^{\infty}dx\,\phi_{k}\left(x\right)\,\bar f_{s}^{3}\left(x\right)=\frac{3\sqrt{3}\,\pi}{32\times2^{3/4}}\,\,.
\label{eq:second order two}
\end{equation}

Substituting equation (\ref{eq:second order one}) in (\ref{eq:asm second order}), one gets
\begin{equation}
\ddot{f}_{r}-f_{r}^{''}+(3\phi_{k}^{2}-1)f_{r}=3\left(\alpha \bar f_{s}-\phi_{k}\bar f_{s}^{2}\right)A^{2}\,\,.
\label{eq:second order three}
\end{equation}

The expression of the amplitude can be written as (see eq.~(\ref{hats}))
\beq
A(t)= \hat A(t) \cos{\left(\sqrt{\frac{3}{2}}t\right)}\,,
\eeq
where $\hat A(t)$ carries the information of the decay of the amplitude, and  at linear order 
 is a constant. The amplitude at $t=0$ is $A(0)=\hat A(0)$.
 
This means that the source term on the right hand side of the Eq. (\ref{eq:second order three}) has double that frequency. 
Manton et al. \cite{Manton:1996ex} were able 
to show that the solution of this equation for the asymptotic form of the radiation can be written as
\begin{equation}
f_{r}\left(x,t\right)=\frac{3\sqrt{3}\,\pi \hat A^{2}}{8\sinh\left(\pi\sqrt{2}\right)}\cos\left[2\sqrt{\frac{3}{2}}\,t-2x-\arctan\left(\sqrt{2}\right)\right]\,.
\label{eq:second order eleven}
\end{equation}

Using this asymptotic solution for the field we can now find the energy flux radiated to infinity by the oscillating 
bound state. Including a factor of $2$ to account for the radiation toward $x\rightarrow-\infty$ and averaging 
over a period, we get the radiated power to be
\begin{equation}
\frac{dE}{dt}=-0.0112909\, \hat A^{4}\,.
\label{eq:second order twelve}
\end{equation}

Finally, the backreaction on the amplitude of the shape mode can be estimated on the grounds of energy 
conservation. It can be shown analytically that the energy of the field configuration
\begin{equation}
\phi\left(x,t\right)=\phi_{k}\left(x\right)+A\left(t\right)\bar f_{s}\left(x\right)
\label{eq:kink plus shape}
\end{equation}
is given by 
\begin{equation}
E\left(t\right) \approx M_{k}+\frac{3}{4}\hat A^{2}\left(t\right)+ ...
\label{eq:kink plus shape energy}
\end{equation}
provided that the amplitude is small.  

Therefore, using energy conservation we obtain
\begin{equation}
\frac{3}{4}\frac{d \hat A^2}{dt}=-0.0112909\,\hat A^{4}\,,
\label{eq:energy conservation shape}
\end{equation}
which yields
\beq
\hat A(t)^{-2} = A(0)^{-2} + 0.0150546 ~ t ~.
\eeq

This is the final result for the long term evolution of the envelope amplitude
of the bound state in Minkowski space. In the main part of the text we will
compare this with our numerical solutions.

\section{Analytical estimate of the amplitude of the bound state at finite temperature}
\label{analitic-projection}

In section \ref{thermal-bath} we have seen that, at low temperatures, the amplitude of the 
bound state in a kink in thermal equilibrium follows a simple relation given by

\beq
\langle \hat A \rangle \propto \theta^{1/2}~,
\eeq
where $\langle \hat A \rangle $ is the average on an ensemble of
realizations at temperature $\theta$.

Here we give a simple derivation of this fact by first computing the 
probability density for the amplitude of the bound state at finite temperature
and obtain from that the average amplitude. In order to do that, we 
will make the assumption that one can directly read the amplitude of the
bound state by projecting the generic thermal state of the perturbation 
on the bound state mode waveform. This is possible due to the 
orthogonality relation of the different perturbation modes we found for the
kink (see the discussion in Sec.~\ref{decay-flat-space}). It is important to note that this neglects the possible non-linear
interaction between the different perturbations modes and the kink itself. This
will limit the validity of our approximation to small amplitude perturbations or
small temperatures.

As we described in Sec.~\ref{phase-transition}, our initial configuration for a thermal state is given by
a Gaussian Random Field with a thermal spectrum given by Eq. (\ref{eq:occupation number tic phys}),
which in the classical limit becomes,
\begin{equation}
\langle |\alpha_{n}|^{2}\rangle\approx\frac{T}{\omega_{n}}\,.
\label{eq:occupation-classical}
\end{equation}
We can now project this configuration over the bound state by following the general
prescription given by Eq. (\ref{numerical-amplitude}). Notice that in our discrete lattice this is nothing
more than a linear combination of the coefficients for $\phi(x,t=0)$.  Similarly,
one can find the initial velocity of the state by the analogous projection in the velocity
field. Doing this exercise, we note that the distribution of the amplitude and the velocity must
be given by another Gaussian Random Field with different values of their variance.

In our numerical experiments we characterized the amplitude of the bound
state by its maximum value in the course of one oscillation, what we have denoted
by the envelope amplitude $\hat A$. This means that
in order to make a meaningful comparison with the calculation of the instantaneous
projection we need to integrate over the possible phases that we happen
to get in the initial conditions. 

Following a somewhat lengthy but straightforward calculation one arrives to
the probability density for the amplitude of the bound state obtained with this
procedure to be
\beq
\rho(\hat A) = {\cal N}(\theta) \hat A~\exp{\left(- \frac{\hat A^2}{\sigma^2 \theta}\right)} I_0\!\!\left( \frac{\hat A^2}{D^2 \theta}\right)
\label{probability-distribution-A}
\eeq
where $I_0$ is the modified Bessel function of the first kind and the
coefficients $\sigma$, $D$ and ${\cal N}$ can be computed in terms
of the parameters of the model and the proper normalization for the 
probability density.

Finally, taking this probability density into account, one can obtain that
\beq
\langle \hat A \rangle \propto \theta^{1/2}~,
\eeq
which is in pretty good agreement with the dependence found in Fig.~\ref{fig:amplitud-versus-T}. Note
that at  temperatures larger than $\theta \sim 0.5$ the numerical results
start to deviate from the present analytical estimate. We attribute this
discrepancy mainly to the fact that at those energies we are outside the
linear regime.

\begin{figure}[h!]
\includegraphics[width=15.6cm]{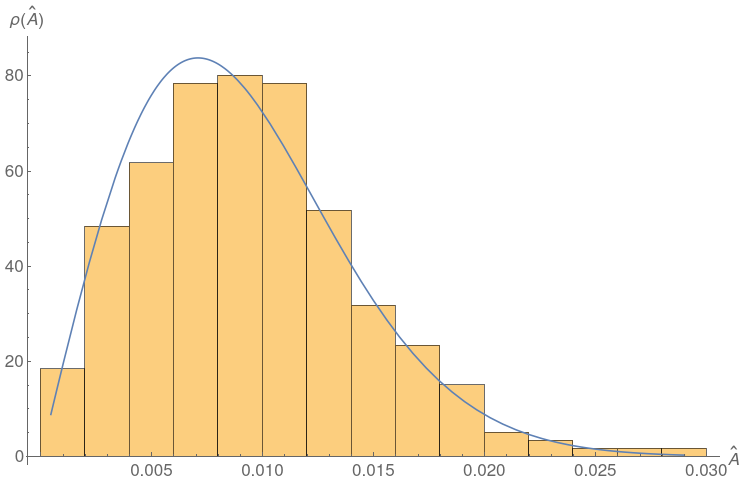}
\caption{\label{bla}Histogram of the probability density for the values of the amplitude of the bound state
$\hat A$ at a temperature $\theta = 10^{-4}$ from 300 simulations. The solid line
represents the estimate of this distribution computed in Eq. (\ref{probability-distribution-A}).}
\end{figure}

Finally, we have tested these results numerically as well by computing the 
distribution of the observed values of the amplitudes of the bound state for 300 realizations
of the initial conditions at a temperature of $\theta = 10^{-4}$. We show a reasonably good agreement in Fig.~\ref{bla} by showing the amplitude distribution of these realizations and its comparison with the
analytic curve given by Eq. (\ref{probability-distribution-A}).

\end{document}